\documentclass[aps,prb,showpacs,twocolumn,superscriptaddress,longbibliography,amsmath,amssymb,floatfix]{revtex4-2}
\usepackage[pdftex]{graphicx}
\usepackage{color}
\usepackage[pdftex,bookmarks,colorlinks,breaklinks]{hyperref} % PDF hyperlinks, with coloured links
\hypersetup{linkcolor=red,citecolor=blue,filecolor=dullmagenta,urlcolor=blue} % colored links
\hypersetup{pdftoolbar=true,pdfpagemode=UseNone,pdfstartview=FitH, colorlinks=true,extension=}
\usepackage{siunitx}
\begin{document}
\title{Experimental Observation of Multifractality in Fibonacci Chains}
\author{Mattis Reisner}
\affiliation{Université Côte d’Azur, CNRS, Institut de Physique de Nice (INPHYNI), France}
\author{Yanel Tahmi}
\affiliation{Université Côte d’Azur, CNRS, Institut de Physique de Nice (INPHYNI), France}
\author{Fr\'{e}d\'{e}ric Pi\'{e}chon}
\affiliation{Laboratoire de Physique des Solides, Universit\'{e} Paris-Saclay, 91400 Orsay, France}
\author{Ulrich Kuhl}
\affiliation{Université Côte d’Azur, CNRS, Institut de Physique de Nice (INPHYNI), France}
\author{Fabrice Mortessagne}
\email{fabrice.mortessagne@univ-cotedazur.fr}
\affiliation{Université Côte d’Azur, CNRS, Institut de Physique de Nice (INPHYNI), France}

\date{\today} % Leave empty to omit a date

\begin{abstract}
The tight-binding model for a chain, where the hopping constants follow a Fibonacci sequence, predicts multifractality in the spectrum and wavefunctions. 
Experimentally, we realize this model by chains of small dielectric resonators with high refractive index ($\epsilon_r \approx 45$) of cylindrical form that exhibit evanescent coupling. 
We show that the fractality of the measured local density of state (LDOS) is best understood when the sites are rearranged according to the similarities in their local surrounding, i.e., their conumbers. 
This allows us to deduce simple recursive construction schemes for the LDOS for the two cases of dominant strong and weak coupling, despite our limited resolution due to non-zero resonance width and size constraints. 
We measure the singularity spectrum and the fractal dimensions of the wavefunctions and find good agreement with theoretical predictions for the multifractality based on a perturbative description in the quasi periodic limit.
\end{abstract}

%\keywords{first keyword, second keyword, third keyword}
\maketitle

\section{Introduction}
The question of understanding wave propagation phenomena in inhomogeneous media transcends almost all types of waves (gravitational, seismic, sound, fluid, electromagnetic, and quantum), ranging from the largest to the smallest wavelength and frequency scales imaginable. Since the fundamental work of P.~W.~Anderson on quantum electrons in disordered systems\cite{Anderson58}, it has been well established that interference effects induced by multiple random elastic scatterings can strongly modify wave propagation in such a way that, depending on the strength of the disorder, three regimes can be distinguished. For a ``weak disorder'', such that the mean free path of the scattering $\ell$ is much larger than the considered wavelength $\lambda$, the waves remain extended and propagate in a diffuse way. For a ``strong disorder'' ($\ell \ll \lambda$), the waves are ``exponentially localized'' in real space and cannot propagate anymore. At the transition between these two regimes, there is a critical regime characterized by a multifractal distribution of wave amplitudes in real space and associated with an anomalous diffusive propagation of wave packets~\cite{weg80,sch91,fast92,sch96,bra96,huc97,evers08}. Several recent experiments have succeeded in revealing such a critical regime with multifractal waves\cite{Morgen03,Hashi08,Faez09,Jack21}.

Beyond the disordered systems at the critical point, many numerical studies have shown that waves propagating in quasi-crystalline structures have generically multifractal properties with the particularity of having tunable fractal dimensions\cite{koh83a,fuji88,fuji89,chh89,tsu91}. Several works have linked these fractal properties of waves to the specific geometrical properties of quasi-periodic lattices (two-dimensional tilings and one-dimensional chains)\cite{koh86,suther86b,niu86,koh87,suther87,tok88,niu90,hol91,pie96,pen95,pen98,rep98,mac16,mac17}. Specifically, although quasicrystalline structures are not periodic, they exhibit long-range orientational and translational order and possess properties of self-similarity and high translational repeatability for domains of all scales. Nevertheless, nearly forty years after the discovery of quasicrystals~\cite{shech84}, there is currently no experiment in real or meta materials that has clearly demonstrated these multifractal properties of waves, even in the simplest and most studied paradigmatic example, i.e., the Fibonacci chain (see the recent review by A. Jagannathan\cite{Jaga21}). However, a recent experiment using cavity polaritons propagating in a Fibonacci chain structure has succeeded in revealing the fractal character of the eigenfrequency spectrum and also in verifying the gap labeling in agreement with theoretical predictions\cite{Tanese14}.

\begin{figure}[b]
    \centering
    \includegraphics[width=\linewidth]{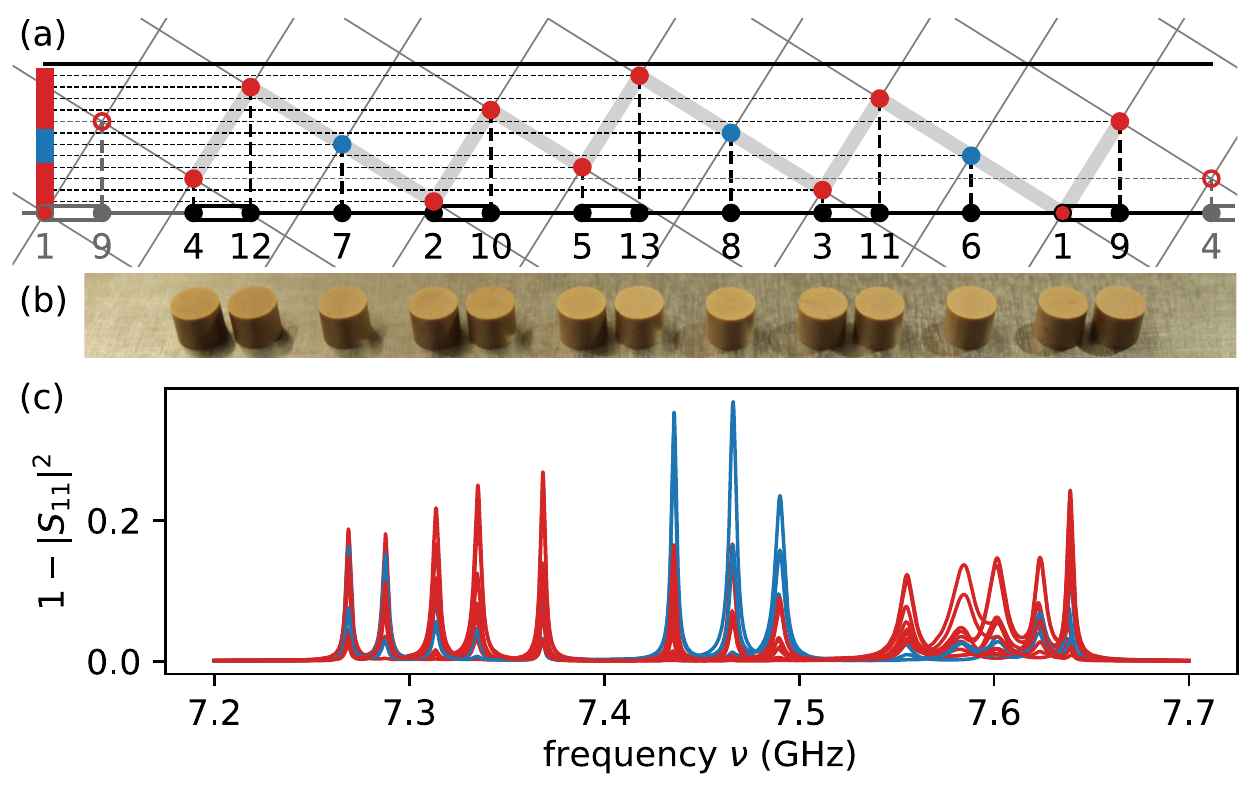}
    \caption{\label{fig:conumbering}
    (a) Example of the cut and project method for the 7th approximate defined by a slope $\omega_7=5/8$ and a motif of $F_{7}=13$ sites. The projection on the horizontal axis dictates the arrangement of the sites of the chain according to strong ( black double-line) and weak (black line) couplings. Each site is reordered according to its local environment on the perpendicular (vertical) axis, the resulting conumber $c(i)$ is indicated under each site at position $i$. Along the perpendicular axis, “atomic” (in blue) and “molecular”  (in red) sites are clustered in 3 groups. 
    (b) Photo of an experimental Fibonacci chain made of 13 resonators. 
    (c) Measured reflection spectra $1-|S_{11}|^2$ for each resonator in the chain shown in (b, where the colors differentiate the atomic (blue) and molecular (red) sites.
    }
\end{figure}

In this work we present, on the one hand, the first experiment that explicitly demonstrates the existence of a simple recursive scheme to reconstruct the fractal properties of the local density of states of the waves on the Fibonacci chain. On the other hand, we quantitatively characterize these multifractal properties and show good agreement between the measured fractal dimensions and those predicted by the simplest modeling of the experiment.

\section{Fibonacci chains of coupled microwave resonators}
For our experimental studies we use a versatile microwave setup that implements a tight-binding system \cite{bel13}. It is based on high index cylindrical dielectric resonators (TiZrNbZnO, Exxelia serie E6000, $n\approx 6.7$, radius $r=\SI{3}{mm}$, height $h=\SI{5}{mm}$) sandwiched between two metallic plates and evanescently coupled. The isolated resonators have a resonance frequency at $\nu_0 \approx \SI{7.45}{GHz}$ with a line width of $\Gamma \approx \SI{2}{MHz}$. The variation of $\nu_0$ between different resonators is within the line width. For further details on the experimental setup and its relevance for topological photonics, see~\cite{rei21}. 

The experimental chains are built following the cut and project method (C\&P). 
The C\&P method can be used to construct all $n$th periodic approximates $C_n$ of the Fibonacci chain, up to its quasiperiodic structure for $n \rightarrow \infty$. It consists in projecting sites in a given interval of a two-dimensional (2D) regular grid onto a line that is cutting the grid with a slope $\omega_n=F_{n-2}/F_{n-1}$, as can be be seen in Fig.~\ref{fig:conumbering}(a). The $F_n$ are the Fibonacci numbers defined via $F_n=F_{n-2} + F_{n-1}$ with $F_1=1$ and $F_2=1$. Note that in the limit $n \rightarrow \infty$, the slope $\omega_n$ tends toward the inverse of the golden ratio $\omega = \omega_{\infty} = \left(\frac{1 + \sqrt{5}}{2}\right)^{-1}$. Due to the irrational nature of $\omega$, the resulting structure is quasiperiodic, whereas for any rational approximation $\omega_n$, the $C_n$ chain exhibits an infinite repetition of the same pattern of $F_n$ sites.

\begin{figure}[t]
    \centering
    \includegraphics[width=.98\linewidth]{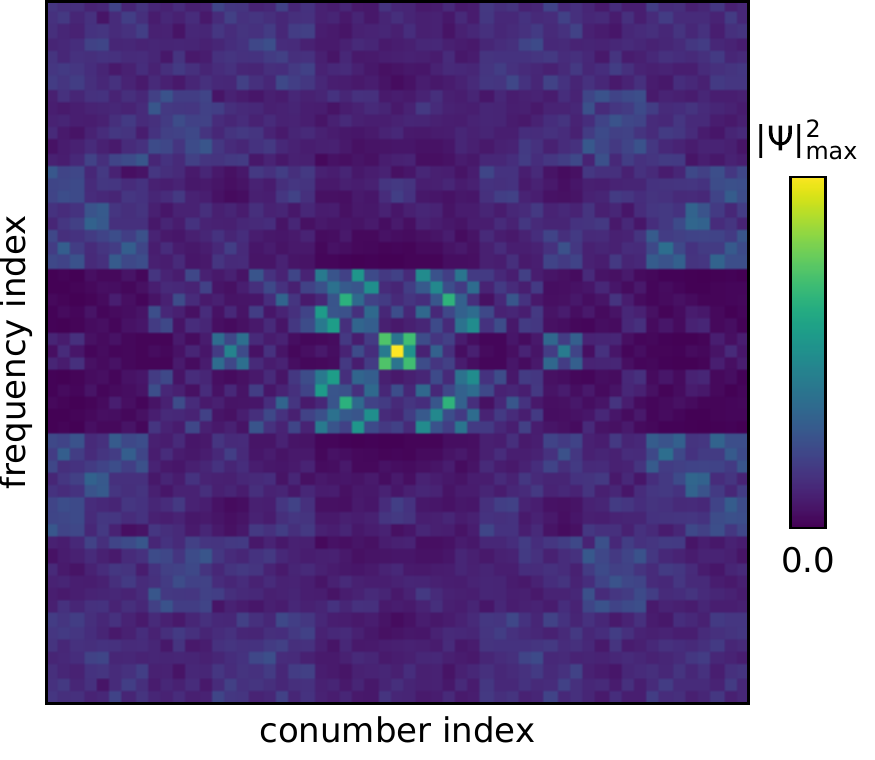}
    \caption{\label{fig:heatmap_55}
    Experimentally extracted LDoS of 55 resonators arranged according to their conumber index $c(i)$ averaged over all 8 permutations.
    }
\end{figure}

The projected points (black points in Fig.~\ref{fig:conumbering}(a)) divide the line in intervals that only have two different lengths $A$ (black line) and $B$ (black double-line). In a sequence of $F_n$ intervals, the ratio between the number of $B$ and $A$ is given by $\omega_n$. Very differently, the sites projected onto the perpendicular axis occupy equally spaced and reordered positions: The sites whose projection on the horizontal axis are surrounded by two $A$ intervals (further referred to as atomic sites) are clustered around the center, whereas those embedded in $ABA$ sequences (further referred as molecular sites) are grouped at the sides -- at the bottom for the sites between $AB$, at the top for $BA$. This way of referring to the sites not by their index $i$ but by their projection on the perpendicular axis [see Fig.~\ref{fig:conumbering}(a)] is called conumbering and was first introduced by R.~Mosseri~\cite{mos88,Sire90}.

From there, different experimental strategies can be followed: either the two letters are associated with two different couplings between resonators, or they are used to account for two different resonant frequencies. We will implement the first one here, thus introducing two coupling, $t_A$ and $t_B$, or, equivalently, two distances $d_A$ and $d_B$. This experimental choice offers two scenarios: either $\rho=t_A/t_B>1$, which corresponds to the dominant strong coupling scenario, or $\rho=t_A/t_B<1$, the dominant weak coupling scenario. The main part of the study reported here will make the use of the second scenario, but we show in Appendix~\ref{sect:interchange} that inverting $\rho$ yields interesting results too.

Fig.~\ref{fig:conumbering}(b) shows the experimental realization of a chain of 13 resonators using the direct pattern created by the C\&P procedure. In this case, the dominant weak coupling regime is implemented. For an experimental reason explained below, the last weak coupling is suppressed. This procedure also has the advantage that the experimental chain reproducing an elementary motif of a $C_n$ Fibonacci-approximation generates $F_n$ collective resonance peaks, as can be seen in the spectra plotted in Fig.~\ref{fig:conumbering}(c), where each spectrum is measured individually by a movable loop antenna placed directly over each resonator~\cite{bel13}. This correspondence between the number of resonators and resonances was expected from the fact that the experiment enters into the scope of a tight-binding model with nearest-neighbor couplings~\cite{bel13}. The spectra measured at molecular site positions are plotted in red, and in blue for atomic sites. It is worth noting that the bunching of sites revealed  by the conumbering procedure has its counterpart in the spectrum. Indeed, one can clearly see that the three resonances within the central band are mainly localized on atomic sites, while the two side bands are dominated by states located at the molecular sites. This correspondence of the frequency index of states and the conumber index of sites arises from the equivalent paths of renormalization that are used to describe band-labels and sites in a perturbative renormalization scheme, when the chains are constructed by a recursive inflation~\cite{Thiem12,mac16}. 

In a first step, the experimental Fibonacci chains we implement are limited to a single repetition of a $F_{n}$-letter motif, with an averaging over different allowed permutations. To reduce finite-size effects, we constrain the experiment to permutations that \emph{(i)} generate patterns whose infinite repetitions $C_n$ would be linked by weak coupling, and \emph{(ii)} impose that the elementary chain ends on both sides by a strong coupling. Each chain is thus made of $F_n$ sites and $F_n-1$ couplings, as illustrated in Fig.~\ref{fig:conumbering}(b) for $F_7=13$. In practice, for a motif of $F_{10}=55$ resonators, in the dominant weak-coupling regime ($\rho<1$), there are 8 different permutations that start and end on a strong coupling. 

We measure the spectrum over each resonator for all permutations for the coupling strengths $t_A=\SI{81}{MHz}$ and $t_B= \SI{126}{MHz}$, corresponding to distances $d_A=\SI{8}{mm}$ and $d_B=\SI{7}{mm}$. The relation between coupling strength $t$ and separation $d$ between two resonators is extracted from two-resonator measurements~\cite{bel13,rei21}. We chose these values in order to have the least possible overlap between resonances in the spectra, while keeping $\rho=t_A/t_{B}=0.64$ reasonably small, for the best visible contrast. The resonance amplitudes $\psi_j(i)$ of each peak $j$ of the measured spectrum above resonator $i$ are extracted via a harmonic inversion method~\cite{Main1999} and a density-based clustering algorithm~\cite{aub20}. Additionally, we symmetrize the results with respect to the central frequency index as the resonance widths for the higher frequency bands are larger, and thus stronger overlapping makes it impossible to extract~\cite{rei21}. Finally, we obtain a discretized form of the local density of states $\text{LDoS}(i,j)=|\psi_j(i)|^2$, where $|\psi_j(i)|^2$ represents the wavefunction intensity of state $j$ evaluated over resonator $i$~\cite{rei21,bel13}. An example spectrum, a detailed description of the data analysis and the LDoS for all configurations can be found in  Appendix~\ref{sect:dataproc}.
Fig.~\ref{fig:heatmap_55} shows the experimentally obtained $\text{LDoS}(c(i),j)$ after normalization, rearrangement of the position index according to the conumbering procedure, and averaging over the 8 permutations. It is exhibiting a fractal structure and a symmetry between frequency index $j$ and conumber index $c(i)$ is clearly visible: The plot is almost invariant under the exchange of the conumber/frequency axis.

\begin{figure}
    \centering
    \includegraphics[width=\linewidth]{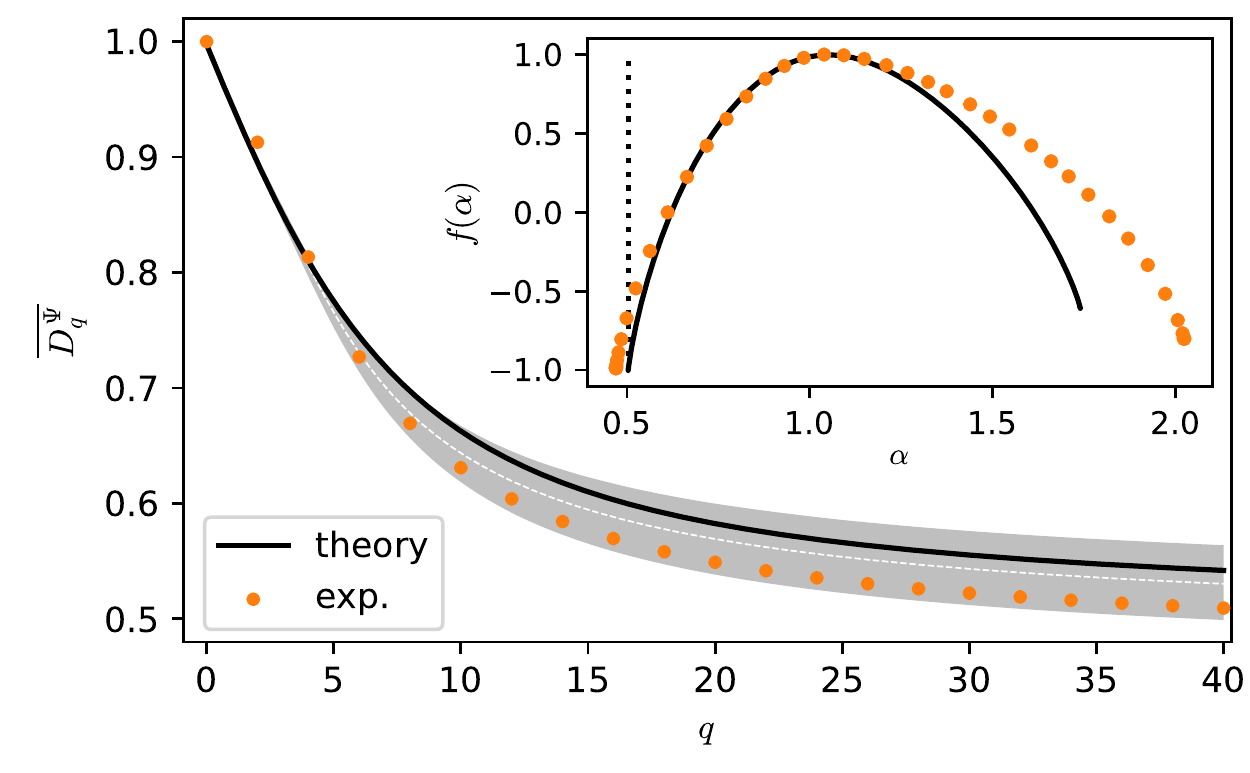}
    \caption{\label{fig:fractal_dimension}
    Spectrally averaged fractal dimension $\overline{D^\psi_q}$ versus the multifractal parameter $q$, experimentally extracted using a box-counting method (orange points) compared to theoretical predictions (solid black line). The gray area highlights the 90\,\% confidence interval obtained from tight-binding simulations (see Appendix~\ref{sect:boxcount}), the dashed line indicates the mean expectation value. 
    The inset shows the theoretical (solid line) and experimental (orange points) spectrally averaged singularity spectrum $f(\alpha)$, where the theoretical $\alpha_\text{min}=\overline{D_{+\infty}^\psi}$ is indicated by a vertical dotted black line.}
\end{figure}

\section{Multifractal dimensions}

A characterization of the multifractal properties of wavefunctions is given by their fractal dimensions $D_q^\psi(j)$ which can be deduced from the scaling,  with the length $F_n$,  of generalized inverse participation numbers~\cite{mac16}:
\begin{equation}\label{eq:individual fractal dimension}
    \chi^{(n)}_q(j) = \sum_i |\psi^{(n)}_j(i)|^{2q}  \underset{n \rightarrow \infty}{\sim} F_n^{-\left(q-1\right) D_q^\psi(j) }\,. 
\end{equation}
The multifractal parameter $q$ allows a selective visualization of the systems at different magnitude scales such that varying $q$ from $-\infty$ to $+\infty$, the dimensions
$D_q^\psi(j)$ decrease from $D_{-\infty}^\psi(j)=\alpha_{max}(j)$ (small intensities) to $D_{+\infty}^\psi(j)=\alpha_{min}(j)$ (large intensities)~\cite{bra96,huc97}.
We further define the frequency-averaged fractal dimension $\overline{D_q^\psi}$ by averaging over all states~\cite{mac16}:
\begin{equation}\label{eq:averaged fractal dimension}
   \left< \chi^n_q(j) \right>_j = \frac{1}{F_n}\sum_j \chi^{(n)}_q(j)  \underset{n \rightarrow \infty}{\sim} F_n^{-\left(q-1\right) \overline{D_q^\psi }}\ .
\end{equation}
As $F_{10}= 55$ is far from the limit $n \rightarrow \infty$, we extract the multifractal dimensions using a box-counting algorithm on the LDoS of Fig.~\ref{fig:heatmap_55}(c)~\cite{thi13} (see Appendix~\ref{sect:boxcount}).

In Fig.~\ref{fig:fractal_dimension} one can see the extracted frequency-averaged fractal dimension $\overline{D_q^\psi}$ as a function of the multifractal parameter $q$ (orange points). We compare it with the frequency-averaged fractal dimension $\overline{D_q^\psi}$ obtained from a theoretical prediction based on a renormalization-group approach, formulated in the limit $\rho \ll 1$ and developed until the order $\rho^{4q}$~\cite{mac16}; in the experiment, $\rho=0.64$. We further estimate a 90\,\% confidence interval for the experiment, by performing tight-binding simulations of the system that account for the variances in the positioning of the resonators and the fluctuations of their resonance frequency. Details on the procedure can be found in the Appendix~\ref{sect:boxcount}. Although far from the strong modulation limit ($\rho \ll 1$), a good agreement between experimental and theoretical values of $\overline{D_q^\psi}$ is obtained, and both curves lie within the estimated confidence interval (see Fig.~\ref{fig:fractal_dimension}). For large $q$, an offset is noticeable between theory and experiment, which could eventually be explained by experimental fluctuations, but even the average value of the simulated $\overline{D_q^\psi}$ (white dashed line) shows an offset. This is mainly due to the finite system size, since the theory was formulated in the quasiperiodic limit (see Appendix~\ref{sect:boxcount}). Note that, although Eq. (\ref{eq:averaged fractal dimension}) is invariant to inverting index $c(i)$ and $j$, our method to calculate $\overline{D_q^\psi}$ via a box counting algorithm is not. Nevertheless, interchanging the conumbering index $c(i)$ with the frequency index $j$ upon the calculation of $\overline{D_q^\psi}$ leads to two hardly distinguishable curves (not shown in Fig.~\ref{fig:fractal_dimension}), further emphasizing the equivalence between conumbers and frequencies.

\section{Singularity spectrum \texorpdfstring{$f(\alpha)$}{f(α)} of the wavefunctions}

An alternative and complementary characterization of the multifractal properties of wavefunctions is given by the so-called  singularity spectrum $f(\alpha)$ \cite{hal86,koh87,fuji89,hol91,pen95,pen98,mac17}. Qualitatively, multifractality encodes the fact that for a given resonance $j$ (resp. for a given resonator $i$) there exists a distribution of anomalous power scaling exponents of the LDoS as a function of the motif length $F_n$: $|\psi_j(i)|^2\propto F_n^{-\alpha}$ with an exponent $\alpha(j,i)$ that depends on $j$ and $i$.
For a plane wave $\alpha(j,i)=1$ therefore when $\alpha<1$ it corresponds to \emph{anomalous} large wavefunction intensities whereas $\alpha>1$ is associated with \emph{anomalous} small intensities. For each exponent $\alpha$ one can also associate a probability $F_n^{f(\alpha)-1}$ to find the exponent $\alpha$ with $0\le f(\alpha) \le 1$. The singularity spectrum $f(\alpha)$ measures the fractal dimensions of interwoven sets of points with different singularity strength $\alpha$.

\begin{figure}
    \centering
    \includegraphics[width=\linewidth]{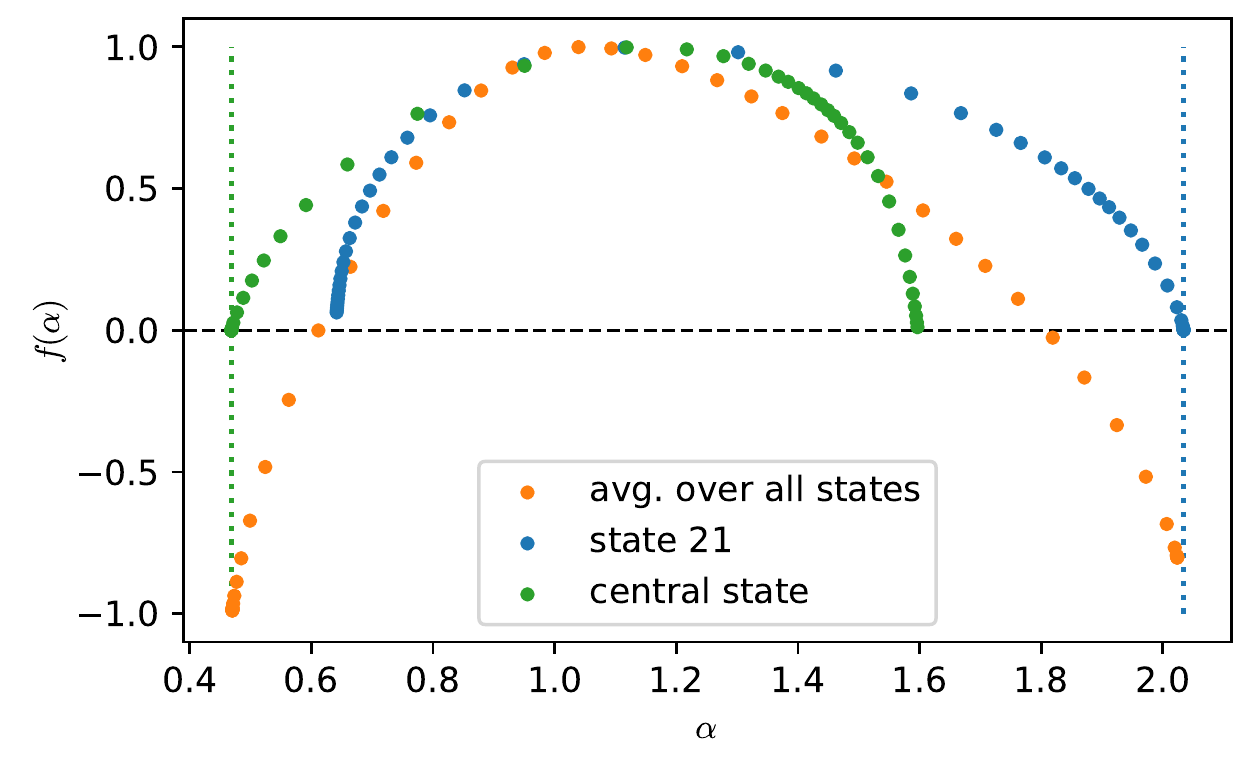}
    \caption{Spectrally averaged singularity spectrum $f(\alpha)$ calculated via a box-counting method (orange points), together with the individual singularity spectra $f_j(\alpha_j)$ of the central state ($j=27$, green points) and the outmost state of the central atomic cluster ($j=21$, blue points). The green dotted line highlights the $\alpha_\text{min}$ of the central state, which also corresponds to the $\alpha_\text{min}$ of the spectrally averaged $f(\alpha)$. The blue dotted line highlights the $\alpha_\text{max}$ of the outmost state of the central atomic cluster, which also corresponds to the $\alpha_\text{max}$ of the spectrally averaged $f(\alpha)$.}
    \label{fig:supp:f(alpha)}
\end{figure}

While the singularity spectrum $f(\alpha)$ can be directly obtained from a Legendre transformation of the fractal Dimension $D_{q}^\psi$,
\begin{equation}
    \alpha(q) = \frac{\text{d}}{\text{d}q}[(q-1)D_q^\psi],
\end{equation}
\begin{equation}
    f(q) = q \alpha(q) - D_q^\psi (q-1),
\end{equation}
we decided to extract it experimentally using an independent box-counting method first proposed by Chhabra \& Jensen \cite{chh89}. If we recall Eq.~(\ref{eq:supp:p}), where we define the spatial distribution of each wavefunction $\psi(i)_j$ by calculating the probability $p_b(\psi_j,L)$ inside box $b$ of size $L$, we can then construct a family of normalized measures, 
\begin{equation}
 \mu_b(q,\psi_j,L)=p_b(\psi_j,L)^q/ \sum_{b=1}^B  p_b(\psi_j,L)^q
 \label{eq:supp:mu}\,.
\end{equation}

From there one can then calculate the Hausdorff dimension of the support of the measure $\mu_b(q,\psi_j)$
\begin{align}
        f_j(q) &=\lim_{L\to 0} \frac{F(q,\psi_j,L)}{\ln L}\\
        &=\lim_{L\to 0}  \frac{\sum_{b=1}^B \mu_b(q,\psi_j,L) \ln(\mu_b(q,\psi_j,L))}{\ln L},
         \label{eq:supp:F}
\end{align}
and the singularity strength 
\begin{align}
        \alpha_j(q) &=\lim_{L\to 0} \frac{A(q,\psi_j,L)}{\ln L}\\
        &=\lim_{L\to 0}  \frac{\sum_{b=1}^B \mu_b(q,\psi_j,L) \ln(p_b(q,\psi_j,L))}{\ln L}.
        \label{eq:supp:A}
\end{align}

The singularity spectrum $f_j(\alpha_j)$ of a state $j$ can be obtained analogously to the determination of $D_{q}^\psi$ via the box-counting method by evaluating the quantities $F(q,\psi_j,L)$ and $A(q,\psi_j,L)$ for different box-sizes $L$ and linearly fitting them against $\ln L$.
In order to calculate the spectrally averaged singularity spectrum $f(\alpha)$ the formalism has to be slightly adapted. Averaging over different wavefunctions represents a supersampling. In order to average over the different states $j$, we replace the sum over the different boxes $\sum_{b=1}^B$ in Eqs.~(\ref{eq:supp:mu}), (\ref{eq:supp:A}) and (\ref{eq:supp:F}) with the double-sum $ 1/F_n \sum^{F_n}_{j=1} \sum_{b=1}^B$.

In Fig.~\ref{fig:supp:f(alpha)} one can see the calculated experimental spectrally averaged singularity spectrum $f(\alpha)$, together with the individual singularity spectra $f_j(\alpha_j)$ of the central state ($j=27$) and the outmost state of the central atomic cluster ($j=21$) calculated for the largest system size of $F_{10}=55$. The minimum value of $\alpha$, $\alpha_{\text{min}}$, can be directly linked to the maximal wavefuntion intensity $\alpha_{\text{min}}= - \log(|\psi_\text{max}|^2)/\log(F_n)$, while the maximum value of $\alpha$, $\alpha_\text{max}$ can be linked to the overall minimum wavefunction intensity $\alpha_{\text{max}}= - \log(|\psi_\text{min}|^2)/\log(F_n)$. In Fig.~\ref{fig:supp:f(alpha)} the green dotted line highlights the $\alpha_\text{min}$ of the central state, which also corresponds to the $\alpha_\text{min}$ of the spectrally averaged $f(\alpha)$, since it is this exact state that contains the overall maximum wavefunction intensity. Similarly the blue dotted line highlights the $\alpha_\text{max}$ of the outmost state of the central atomic cluster, which also corresponds to the $\alpha_\text{max}$ of the spectrally averaged $f(\alpha)$, since this state contains the overall minimal wavefunction intensity. 

One further notices, that the spectrally averaged singularity spectrum $f(\alpha)$ takes negative values, while the $f_j(\alpha_j)$ of the individual wave functions stays strictly positive. This arises from the supersampling that we perform by averaging over all different states. Regions with negative $f(\alpha)$ correspond to rare wavefunction intensity values that are encountered only for a few states. In this context of supersampling it is actually the relative frequency with which certain wavefunction intensities are appearing, instead of their absolute number, that scales as $F_n^{f(\alpha)}$.

For a more quantitative comparison with theoretical prediction~\cite{Rudi98}, the averaged $f(\alpha)$ obtained by averaging over all $j$ is illustrated in the inset of Fig.~\ref{fig:fractal_dimension}.
For $\alpha\le 1$ (\emph{large} wavefunction intensities) there is a good agreement between the theoretical and the experimental $f(\alpha)$. For larger values of $\alpha>1$ there is a growing disagreement between the experimental and theoretical curves. The latter can be explained by the fact that the region $\alpha>1$ is associated to the smallest intensities, which are necessarily less accurately measured and sampled because they might be below the experimental noise level. 
Note also that because of the averaging procedure the averaged $f(\alpha)$ is no longer positive definite, and it takes negative values \cite{man90,man91}. This \emph{super-sampling} effect has already been observed at the critical point of Anderson transitions \cite{mir00,evers08}.

\begin{figure*}[t]
    \centering
    \includegraphics[width=1\linewidth]{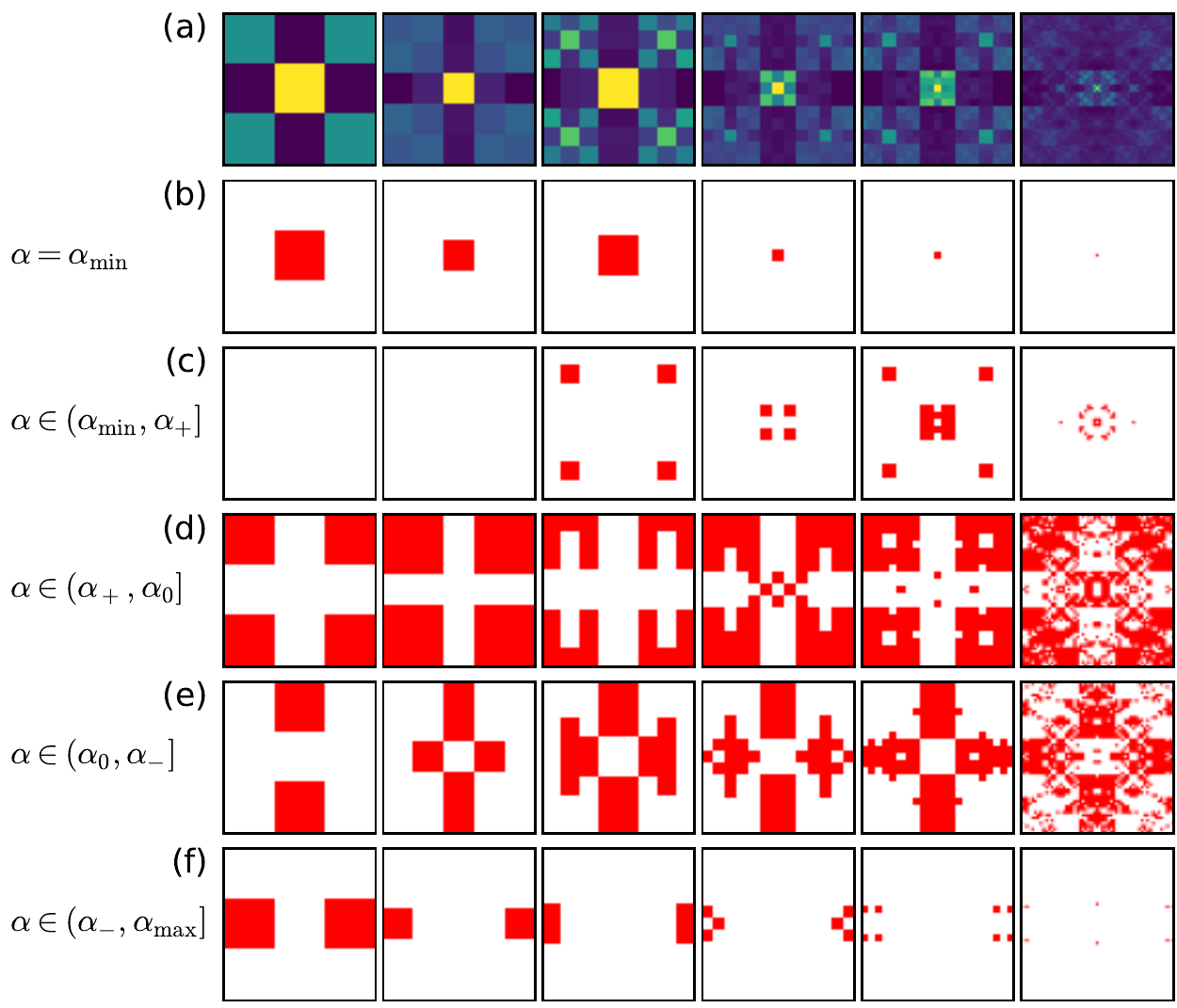}
    \caption{\label{fig:supp:different_sets_alpha}
    (a) Symmetrized $\text{LDoS}(c(i),j)$ for the different approximants, where the approximants are ordered from left to right according to their motif-length $F_n$: $F_4=3$, $F_5=5$, $F_6=8$, $F_7=13$, $F_8=21$, and $F_{10}=55$. 
    The vertical axis of all subplots corresponds to the frequency index $j$, while the horizontal axis corresponds to the conumbering index $c(i)$. 
    (b) - (f) highlighted positions of the pixels in the corresponding LDoS in red, whose intensities lay within the indicated $\alpha$-intervals. 
}
\end{figure*}

To further give a more graphical representation of the singularity spectrum, in Fig.~\ref{fig:supp:different_sets_alpha}, we plot different sets of points whose intensities lay within different intervals defined by certain $\alpha$-values for different approximants obtained from the experimentally extracted symmetrized LDoS, which is presented in the top row. 
Only for this figure we have additionally symmetrized the discretized $\text{LDoS}(j,c(i))$, as the resulting sets of points presented in the 5 lower rows are less affected by experimental fluctuations. The symmetrized LDoS is obtained by independently symmetrizing along both the frequency and conumbering axes.  

Due to finite size effects, the same $f(\alpha)$ does not exist for different approximants. Thus we separate the $f(\alpha)$ into 5 distinct intervals and present their support in the 5 lower rows. 
The borders of the intervals are defined by characteristic points of the $f(\alpha)$ curve to define the intervals. In addition to using $\alpha_\mathrm{min}$ and $\alpha_\mathrm{max}$, we use the point $\alpha_0$, where $f(\alpha_0)=1$, associated with $q=0$ and the two roots of $f(\alpha)$, $\alpha_{+}$ and $\alpha_{-}$,
where $\alpha_{+}$ is the left root, that lays in the region associated with positive values of $q$ and $\alpha_{-}$ is the right root, that lays in the region associated with negative values of $q$ \cite{evers08}.

Figure~\ref{fig:supp:different_sets_alpha}(b) highlights the pixels of the discretized LDoS, which correspond to $\alpha_{\text{min}}$, which are the points that have the maximum wavefunction intensities.
Apart from $F_n=8$, there exists only one pixel with the maximum intensity in the LDoS, which is the central one.
This therefore directly results in its fractal dimension of $f(\alpha_{\text{min}})=-1$, since it is encountered with a relative frequency of $1/F_n$.
 As $F_n=8$ is the only even system size presented in Fig.~\ref{fig:supp:different_sets_alpha}, it does not have a central pixel thus illuminating the central 4 pixels. 
This odd/even difference was already discussed in the previous section (see Fig.~\ref{fig:supp:offset_dimension}), and it leads here to additional oscillations when approaching the limiting value.

In Fig.~\ref{fig:supp:different_sets_alpha}(c) and (f) associated with the interval $\alpha\in (\alpha_\mathrm{min}, \alpha_+]$) and $\alpha\in (\alpha_-,\alpha_\mathrm{min}]$, where $f(\alpha)<0$, the subsets are containing a few states, whereas for (d) and (e) corresponding to $f(\alpha)>0$ the majority of states contribute.

\section{Self similarity}
Another aspect often associated with fractality is the self similarity of structures~\cite{mac17,rep98,koh87,koh86,suther86b,suther87,tok88}. Similar to the recursive construction of the Fibonacci numbers $F_n$, the complete LDoS can be constructed recursively. The procedure is based on the renormalization of atomic and molecular sites~\cite{mac16}
\begin{align}
    |\psi^{(n)}_j(c_i)|^2 &= \overline{\lambda} \cdot |\psi^{(n-3)}_{j'} (c_{i'}) |^2 \quad \text{if $j$ is atomic,}  \label{eq:renormalization_atomic}\\
    |\psi^{(n)}_j(c_i)|^2 &= \lambda \cdot |\psi^{(n-2)}_{j'} (c_{i'})|^2 \quad \text{if $j$ is molecular,}   \label{eq:renormalization_molecular}
\end{align}
where $\overline{\lambda}$ and $\lambda$ are renormalization factors that depend on $\rho$.

\begin{figure}[t]
    \centering
    \includegraphics[width=\linewidth]{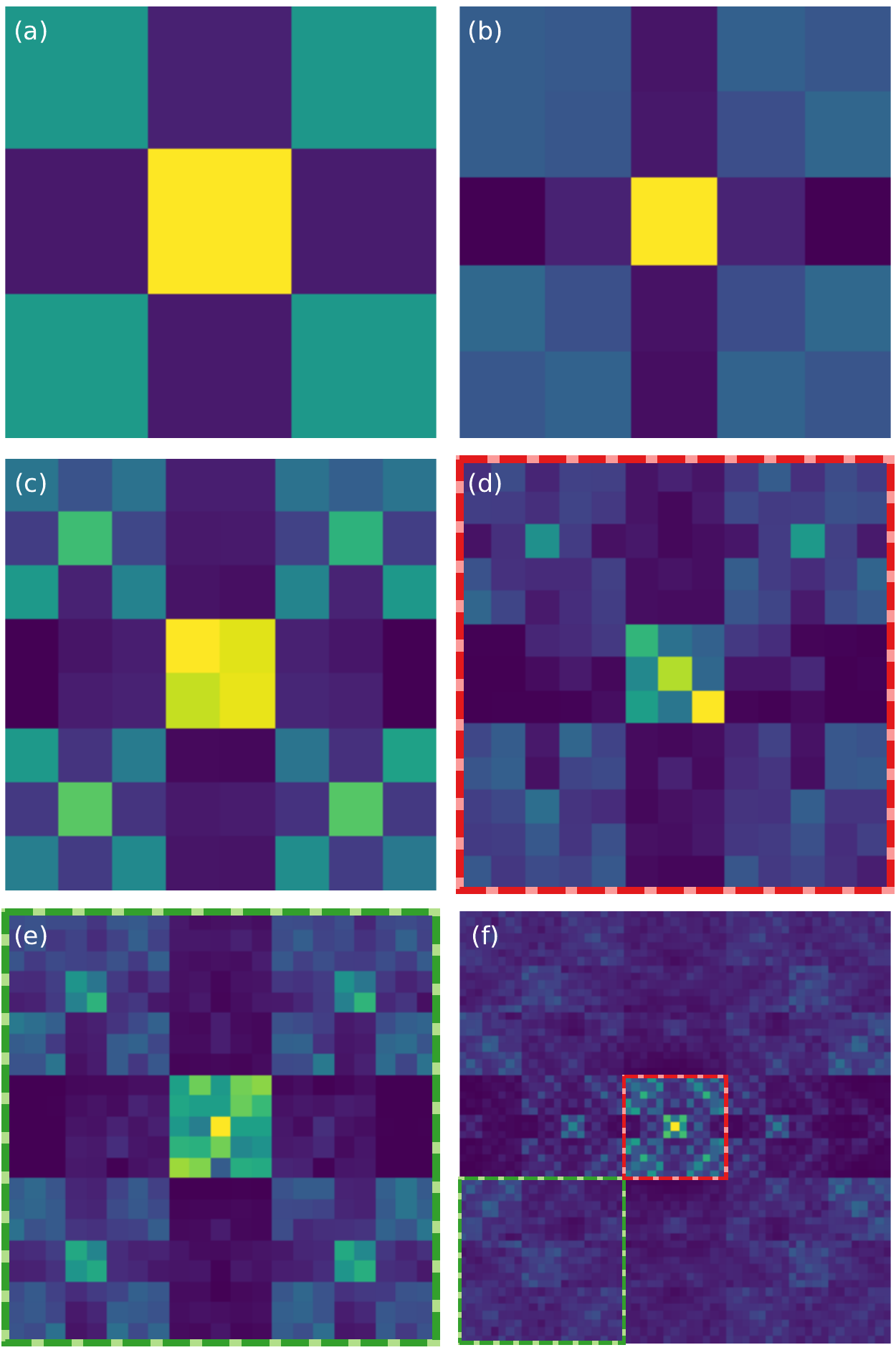}
    \caption{\label{fig:all_heatmaps}
    Conumber-averaged LDoS for different motif length $F_n$: (a) $F_4=3$, (b) $F_5=5$, (c) $F_6=8$, (d) $F_7=13$, (e) $F_8=21$ and (f) $F_{10}=55$ (already presented in Fig.~\ref{fig:heatmap_55}). For all plots the horizontal axis corresponds to the conumber index $c(i)$ and the vertical axis to the frequency index $j$, and the same colormap as in Fig.~\ref{fig:heatmap_55} is used. The green and red squares highlight the recursive construction.}
\end{figure}  

We investigate this recursive construction by experimentally realizing the first periodic approximations (i.e. $F_n=3,5,8,13,21$). Instead of using different permutations of the periodic motif, as we have done previously, we use circular chains, where the basic motif with $F_n$ sites is repeated $N_p$ times. The number of repetitions $N_p$ is chosen such that a ring of around 100 resonators is built for each $F_n$-motif. In this way, the $F_n$ bands, expected for an infinite chain $C_n$, are each populated with $N_p$ states, in contrast with the previous experiment, where a single state was defining the band position. Due to the higher density of states inside the bands, an individual extraction of resonance is not possible anymore and we extracted the LDoS directly from the reflection spectra, where we reduced the weaker coupling to $t_A=\SI{55}{MHz}$ and we enhanced the stronger coupling to  $t_B=\SI{148}{MHz}$, in order to obtain better isolated bands (for details see Appendix~\ref{sect:dataproc}). This allows us to average over equivalent sites and states.

In Fig.~\ref{fig:all_heatmaps} we present the LDoS for the first approximates. Highlighted for $F_{10}=55,F_8=21$ and $F_7=13$ (Fig.~\ref{fig:all_heatmaps}(f), (e) and (d), respectively], the central square (marked in red), which gathers atomic sites and their corresponding states, of the LDoS at order $n$ resembles the complete LDoS of order $n-3$, and the four squares in the corners (molecular sites and frequencies, one marked in green) of the LDoS at order $n$ resemble the complete LDoS of order $n-2$. The recursive construction is also well visible for smaller $n$. We calculate the renormalization factor $\overline{\lambda}$ by integrating the central square (corresponding to atomic sites and states) and $\lambda$ by integrating and averaging over the four corner squares (corresponding to molecular sites and states) in Fig.~\ref{fig:all_heatmaps}(f). We find $\overline{\lambda} = 0.51 $ and $\lambda = 0.42$, which are in reasonably good agreement with theoretical predictions for the quasiperiodic limit $\overline{\lambda}_{\text{theo}} = 0.48 $ and 
$\lambda_{\text{theo}} = 0.43$ for $\rho=0.64$. Further information about the theoretical predictions and the experimental estimation of the renormalization factors can be found in  Appendix~\ref{sect:scaling}.

\section{Conclusion}
In this article, we have shown that the multifractal properties of waves propagating on a quasiperiodic lattice can be unambiguously observed in our finite-size experimental set-up made of coupled dielectric resonators. Our measurements were successfully analyzed using a renormalization group approach. The robustness of the fractality observed will be challenged in the near future by introducing controlled disorders in the experiment~\cite{Mous21}: either a coupling disorder, induced by a small variation of distances between microwave resonators, or a phason disorder, resulting from a local inversion of short and strong bond and giving birth to configuration that cannot be obtained by permutation. Our microwave experimental platform is also well suited to the study of 2D lattices~\cite{rei21}, and it has  already been used to provide new physical insights into the behavior of waves on a Penrose tiling~\cite{Vigno16}. Due to the physical couplings that are not constrained to the edges of the tiles, the tight-binding model implemented in the lattice is not the one usually theoretically and numerically studied. Thus by implementing 2D tiling of codimension 1~\cite{Sire90} in our experiment, as for example the Rauzy tiling~\cite{Vidal00}, one can expect to exhibit richer multifractal properties.

\bibliography{bibtex}
%\newpage
\appendix

%%%%%%%%%%%%%%%%%%%%%%%%%%%%%%%%%%%%%%%%%%%%%%%%%%%%%%%%%%%
\section{Extracting the local density of states from the measured spectra}
\label{sect:dataproc}
A general presentation of our versatile tight-binding microwave experiment can be found in \cite{bel13,rei21}. In the following section, we briefly point out the link between the measured reflection spectrum, the local density of states and the eigenvectors of the tight-binding-system. We consider a tight-binding Hamiltonian $\mathcal{H}_\textrm{TB}$ that describes a system of $N$ coupled resonators, with associated eigenvalues $\{\nu_j\}$ and eigenvectors $\{ c^j \}$ that are used to describe the wavefunctions $\psi_j(\vec{r})=\sum^N_i c^j_i \cdot \psi_0(\vec{r}-\vec{r_i})=\sum^N_i \psi_n(i)$ of the tight-binding system, where $\psi_0(\vec{r})$ is the single resonator wavefunction and $\vec{r_i}$ is the position of the resonators. Assuming a Breit-Wigner form of the scattering matrix and a constant antenna coupling $\sigma$ throughout the whole frequency range, the reflection spectrum is then given by
\begin{equation}
 S(\vec{r},\nu) = 1- \text{i} \sigma \sum^N_j \frac{ | \psi_j(\vec{r})|^2 }{\nu - \nu_j + \text{i} \Gamma_j},
\end{equation}
where $\Gamma_j$ is the decay rate associated with state $j$ and $\vec{r}$ is the position of the measuring antenna. 
One can then derive the local density of states 
\begin{equation}
\rho(\vec{r},\nu)= \frac{1}{\pi \sigma}[1-\Re S(\vec{r},\nu)] = \sum_j |\Psi_j(\vec{r})|^2 \cdot f_{\nu_j,\Gamma_j}(\nu),
\end{equation}
where $f_{\nu_j,\Gamma_j}(\nu)$ are normalized Cauchy distributions around $\nu_j$ with width $\Gamma_j$ ($\int_{-\infty}^{+\infty} f_{\nu_j,\Gamma_j}(\nu) d\nu=1$). 
Due to the typical linewidths $\Gamma$ of a few MHz, in our case of large $N$ the resonance peaks in the spectrum, and thus in the density of states, are strongly overlapping. We therefore define a discretized version of the local density of states in the frequency as well in the space domain 
\begin{equation}
\text{LDoS}(i,j)=|\Psi_j(i)|^2=|\Psi_j(\vec{r_i})|^2 = \lim_{\{\Gamma_j\}\to 0} \rho(\vec{r_i},\nu_j).
\end{equation}
This quantity is evaluated from the reflection spectra, by extracting all resonance amplitudes measured exactly over the center of each resonator. The resonance-amplitudes are normalized so that $\sum |\psi_j(i)|^2=1$, and they can be directly associated with the squared eigenvectors $|c_i^j|^2$ of the tight-binding Hamiltonian $\mathcal{H}_\textrm{TB}$ .

For small system sizes ($N \lesssim 10$), direct fits of the spectra with a sum of complex Lorentz lines can be implemented\cite{rei21}. For larger systems, the overlap between resonances becomes too strong, making the fitting strategy impractical. To extract the $\text{LDoS}(i,j)$, we thus developed two different techniques depending on whether we work with the linear chains of 55 resonators or the circular chains of around 100 resonators.

\begin{figure}
 \centering
 \includegraphics[width=\linewidth]{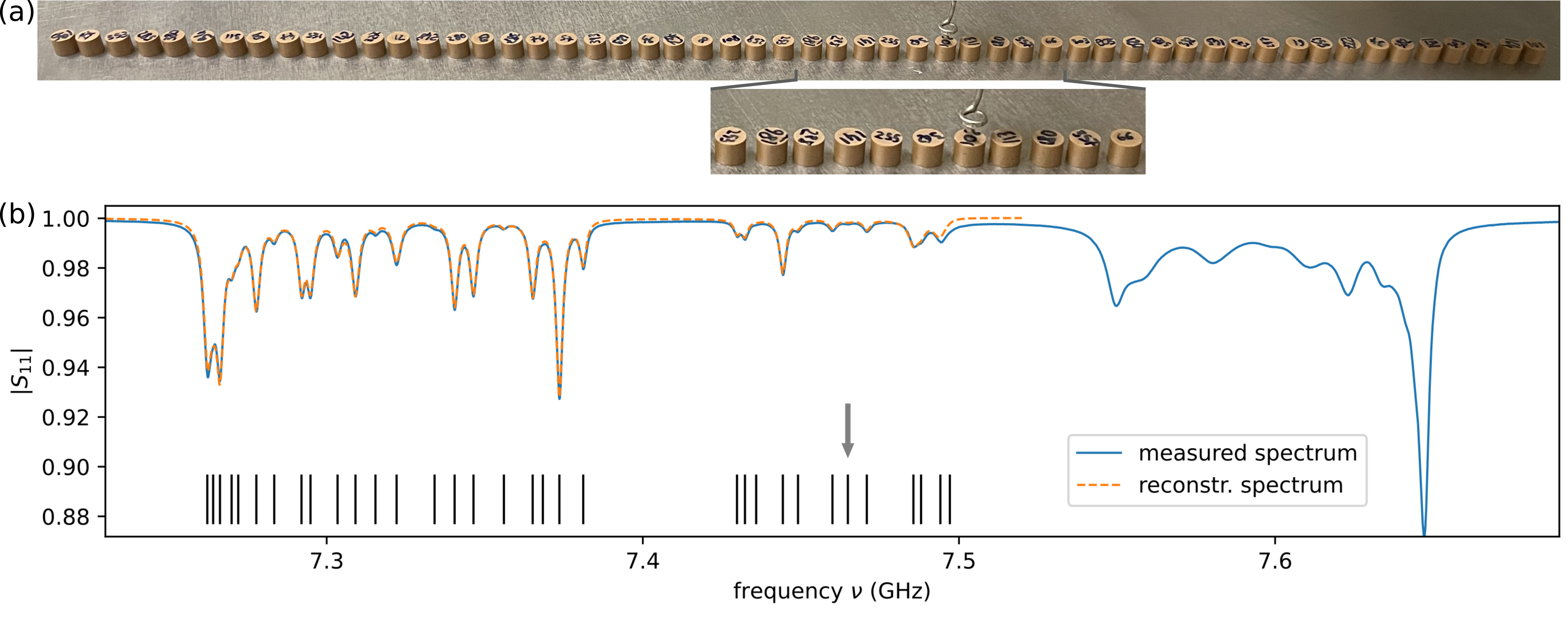}
 \caption{\label{fig:supp:reconstr.}
 (a) picture of the experimental chain of one permutation for the case of dominant weak coupling ($\rho = t_A/t_B < 1$). The metallic top plate that is normally placed above the resonators in order to reduce the system to two dimensions was removed in order to take the picture. Above the resonator at position 33 (counted from the left) we position the loop antenna through which we measure the reflection spectra. (b) reflection spectrum measured at position 33 and the reconstructed spectrum using the resonances obtained via the harmonic inversion method. The black vertical lines mark the extracted resonance positions, and the gray arrow marks the central state ($j=28$), above which we symmetrize the $\text{LDoS}(i,j)$ in order to span the whole frequency range.
 }
\end{figure}

\subsection{Harmonic Inversion and clustering algorithm}

The linear chains are comprised of 55 resonators. A picture of one configuration, with dominant weak coupling, can be seen in Fig.~\ref{fig:supp:reconstr.} (a). The resonators are numerated, in order to identify them and choose only resonators whose resonance frequencies are very close to each other. For the 55 resonators that we use, the difference between the highest and lowest frequencies is around $\SI{3}{MHz}$, the same order of the single resonator linewidth $\Gamma_0$. The spectrum measured above the center of the resonator at position $i=33$ of the chain can be seen in Fig.~\ref{fig:supp:reconstr.} (b), where one can clearly see the overlapping between resonance peaks.

To extract all resonance-amplitudes for each configuration, we use a method based on an algorithm called Harmonic Inversion \cite{Main1999,kuh08}. It is based on the fact that in the time domain, a complex Lorentz line gives rise to an exponential function. Supposing that the time signal (discrete signal with 2$N$ points) only consists of $N$ exponential functions with different complex amplitude and exponents, one can establish a set of nonlinear equations in order to determine all of their parameters. Since the harmonic inversion tries to describe the whole spectrum with a sum of Lorentzian functions, we first have to filter out resonances induced by the non flat baseline of the reflection measurements and by the small fluctuations due to noise. An efficient filtering is obtained by keeping only the resonances whose widths and amplitudes are within a given interval. We then perform a clustering in order to follow each resonance from one antenna-position to the other, regrouping them and associating them with a certain state \cite{aub20}. For each configuration, we adjust the parameters of the density-based clustering algorithm, so that we use the same parameters for all antenna-positions, to avoid manually clustering/adjusting states according to our expectations. 

Figure~\ref{fig:supp:reconstr.} shows the partial reconstruction of the spectrum using the harmonic inversion algorithm, the black horizontal lines indicating the frequencies of the extracted resonances. The quality of the fit is excellent. We limit the reconstructed spectrum to the lower and central frequency band only, since, as described in \cite{rei21}, the higher frequency states have generally greater resonance widths due to different effective antenna-couplings and larger ohmic losses. If only next-nearest-neighbor couplings are present, the system has a CT-symmetry \cite{rehe20} imposing that the spectrum is symmetric around the eigenfrequency of a single resonator. In our experiment we have a next-nearest-neighbour coupling of the order of only 5\,\% of the nearest-neighbour coupling. As a consequence, the latter symmetry is almost preserved. We thus restrict our analysis to the first 28 states (the 28th state is the central state and is indicated by an gray arrow in Fig.~\ref{fig:supp:reconstr.}) and symmetrize the result to expand over the higher-frequency states. Theoretically the eigenvectors of the tight-binding Hamiltonian are normalized in both directions ($\sum_i |c_i^j|^2=\sum_j |c_i^j|^2=1$); the experimentally extracted $\text{LDoS}(i, j)$ should then also be normalized along both the frequency and position axis ($\sum_i \text{LDoS}(i, j)=\sum_j \text{LDoS}(i, j)=1$). Since the antenna-coupling $\sigma$ is slightly dependent on the frequency, and the single-resonance wavefunctions are overlapping \cite{rei21}, the sum of the raw resonance amplitudes over all positions (states) varies about 10\,\% for different states (positions). We thus normalize the extracted wavefunction intensities in both dimensions by alternatingly normalizing them along one direction and then the other, until the difference in normalization along both directions is of the order of $10^{-6}$. We then consider that the extracted $\text{LDoS}(i, j)$ is properly normalized along the two dimensions, which is especially important for the calculation of the fractal dimensions.

\begin{figure*}
    \centering
    \includegraphics[width=\textwidth]{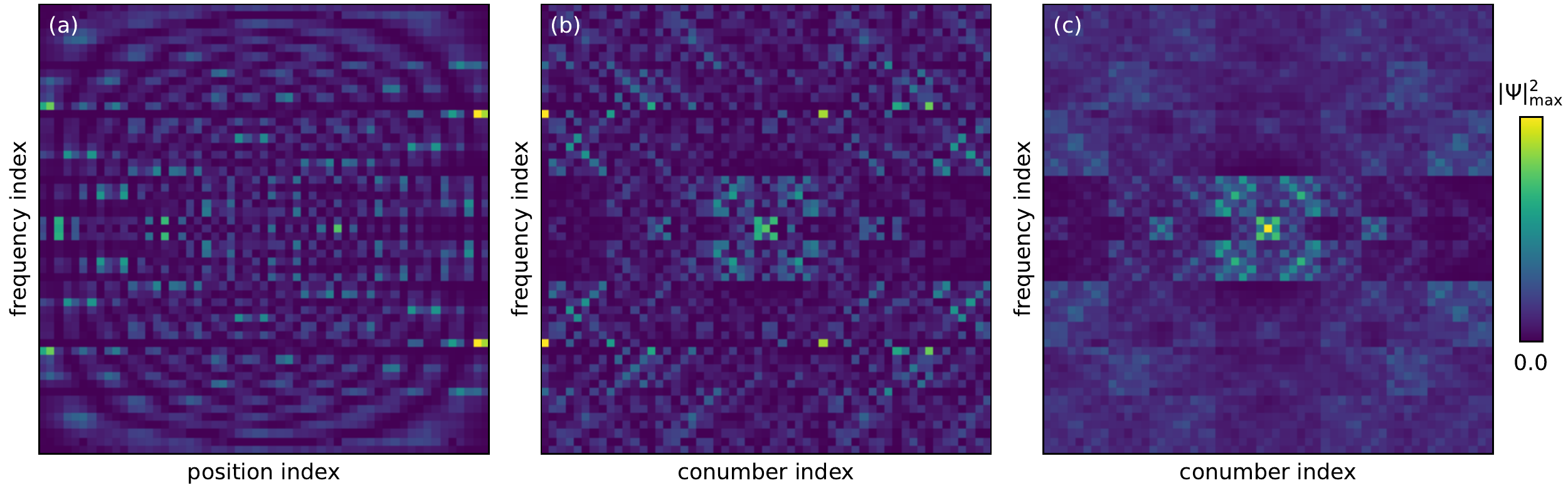}
    \caption{\label{fig:supp:heatmap_55s}
    Experimentally extracted LDoS of a single configuration of 55 resonators arranged according to their position index $i$ (a), rearranged according to the conumber index $c(i)$ (b) and the average over all 8 permutations (c).
    }
\end{figure*}

Figure~\ref{fig:supp:heatmap_55s}(a) shows the local density of states $\text{LDoS}(i,j)$ extracted and normalized according to the procedure described above for a single configuration of a chain made with 55 resonators. Figure~\ref{fig:supp:heatmap_55s}(b) shows a rearranging of the LDoS according to their conumber: $\text{LDoS}(c(i),j)$, and (c) the average over the eight permutations identified in this situation of dominant weak coupling. In Fig.~\ref{fig:supp:heatmap_55s}(a), the LDoSs exhibit typical standing-wave interference patterns due to the finite-size of the chain~\cite{Dutreix2021}, but no hierarchical structure is visible. Reordering the LDoS based on the conumber index provides insight into fractal structures, which are completely revealed by the average over all permutations [see Fig.~\ref{fig:supp:heatmap_55s}(c)].

\subsection{Averaging each frequency-band of circular chains}

\begin{figure}[t]
 \centering
 \includegraphics[width=\linewidth]{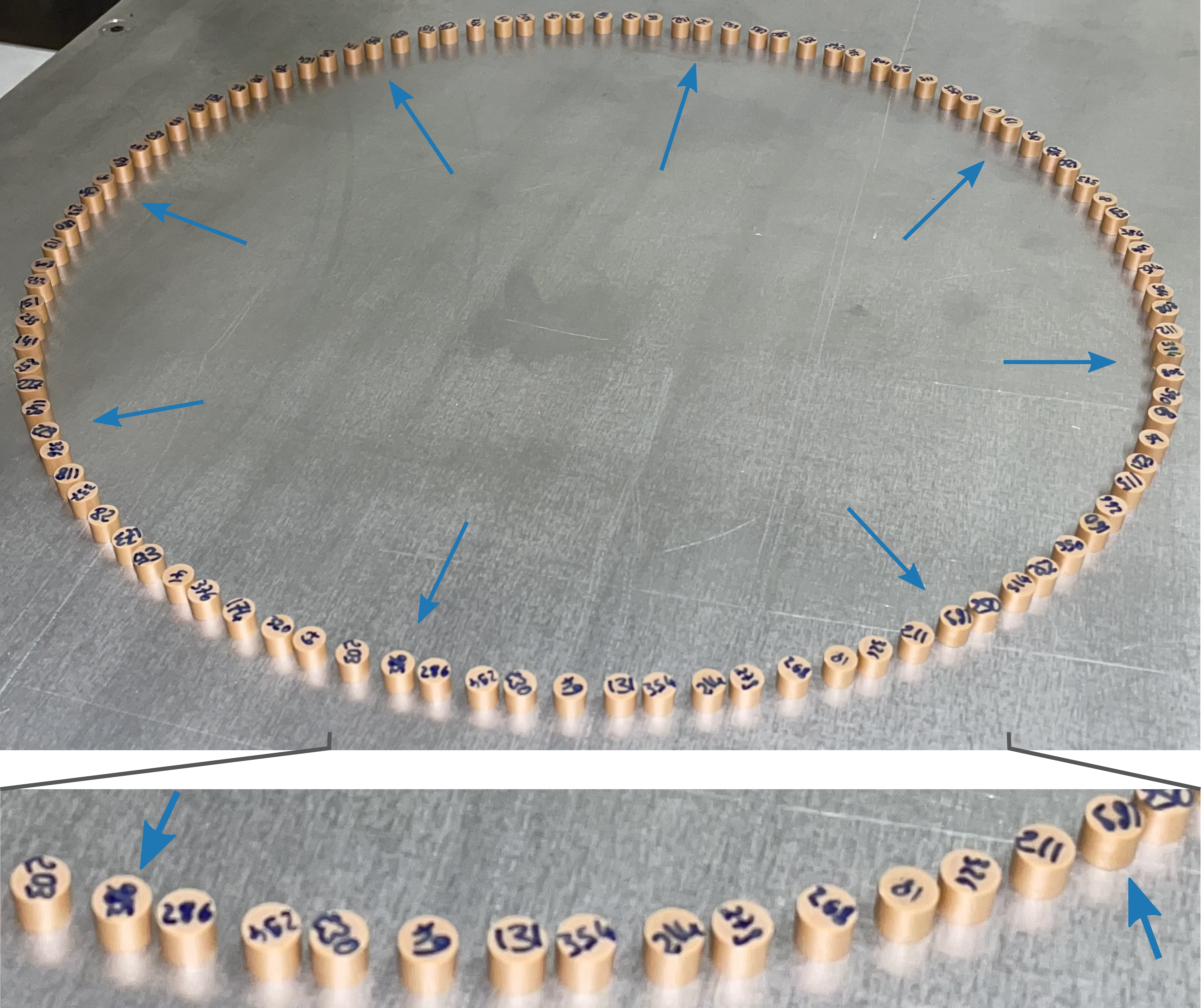}
 \caption{\label{fig:supp:photo_circ}
 (top) Photo of one circular chain, where the basic motif with length $F_n=13$ is repeated $N_p=8$ times, resulting in a total of 104 resonators. To emphasize the periodicity the first resonator of each repeated motif is marked with a blue arrow. (bottom) Zoomed in photo of one motif, where one can identify the ``molecules" (dimers) and ``atoms" (single resonator) that make up the chain.
 }
\end{figure}

The circular chains are made up of smaller motifs (i.e. $F_n=3,5,8,13,21$) that we repeat $N_p$ times while imposing a weak coupling between two consecutive repetitions. The number of iterations $N_p$ is chosen such that a ring of around 100 resonators is built for each $F_n$-motif. In this way, the $F_n$ frequency bands expected for an infinite chain $C_n$ are each populated with $N_p$ states, and they can be individually identified in each reflection spectrum $S_{11}(i,\nu)$ measured over each resonator $i$. In Fig.~\ref{fig:supp:photo_circ}, one can see a photo of the circular chain of resonators for a motif length of $F_n=13$, that was repeated 8 times. Compared to the linear chains where the lowest overlap was sought to identify each state individually, for the circular chains we really want to create $F_n$ energy bands as dense as possible. We therefore enhanced the stronger coupling to $t_B=\SI{148}{MHz}$ and reduced the weaker coupling to $t_A=\SI{55}{MHz}$, in order to obtain better isolated bands. Experimentally, this was done by increasing the longer distance $d_A$ to $\SI{9}{mm}$ (we keep the shorter distance at $\SI{7}{mm}$) and by reducing the distance between the two metallic plates that sandwich the resonators from $\approx\SI{12}{mm}$ to $\SI{8}{mm}$. This alters the evanescent decay of the electromagnetic fields outside of the resonators.

Rearranging the sum over the different states, one can rewrite
\begin{equation}
 \rho(\vec{r_i},\nu)=\sum_{j=1}^{F_n} \sum_{p=1}^{N_p} f_{\nu_{j,p}, \Gamma_{j,p}}(\nu)\cdot |\psi_{j,p}(i)|^2\,,
\end{equation}
where $\nu_{j,p}$ and $\Gamma_{j,p}$ are the resonance frequency and resonance width of the $p$th state within the $j$th frequency band and $|\psi_{j,p}(i)|^2$ is the corresponding wavefunction intensity measured over resonator $i$.

Supposing that the bands are sufficiently isolated, by integrating each frequency band $j$ individually, one can then find
\begin{equation}
 \text{LDoS}(i,j)\propto \int_{\text{band $j$}} \left[1 - \Re{S_{11}(i,\nu)}\right] \,\text{d}\nu\,,
\end{equation}
where we can further average over all indices $i$ that have the same conumbering $c(i)$. 

\begin{figure}
 \centering
 \includegraphics[width=\linewidth]{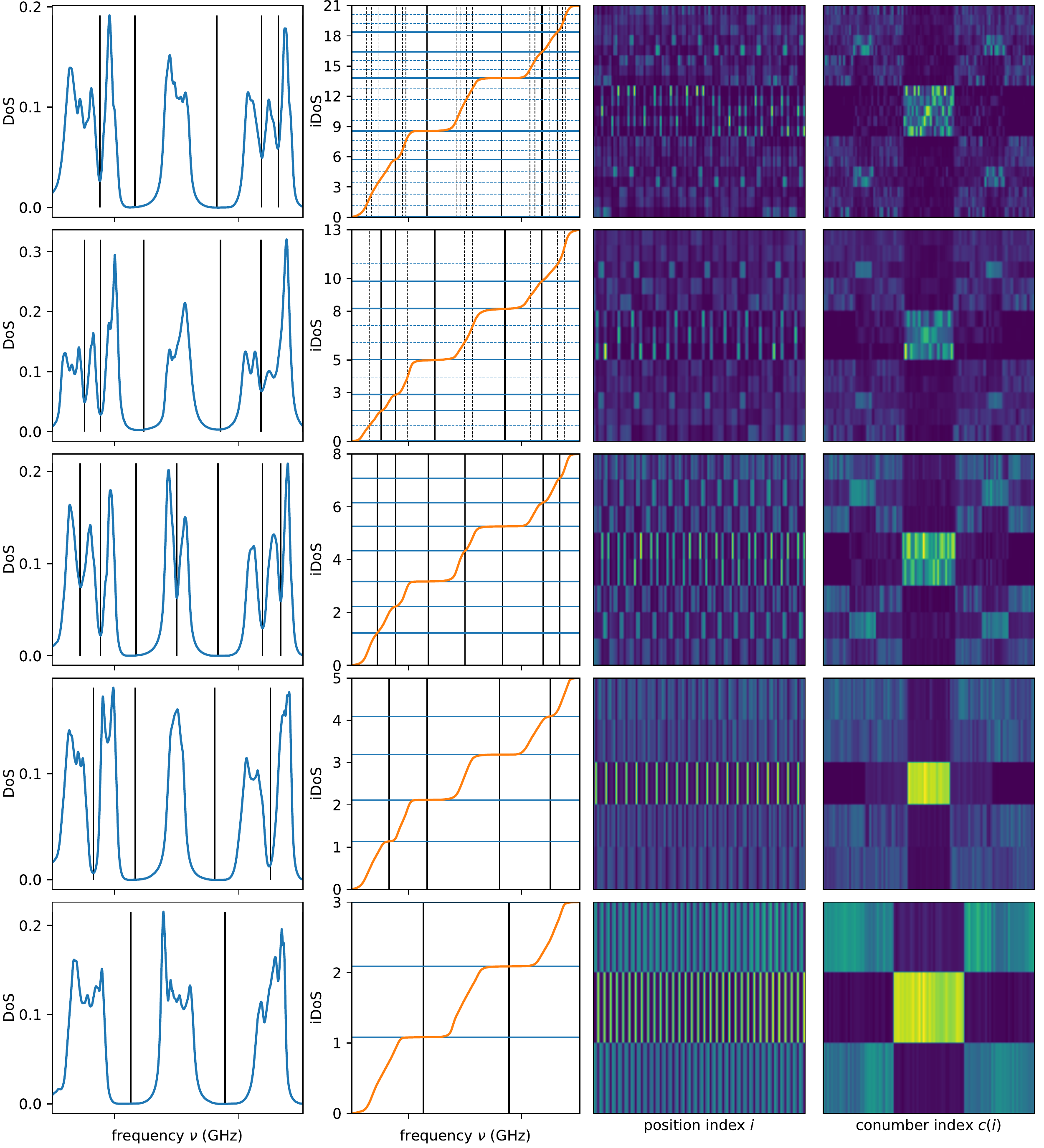}
 \caption{\label{fig:supp:datatreat_circular}
 From the top to the bottom: The different steps of the data treatment procedure for the circular chains with (first row) $F_n=21$, (second row) $F_n=13$, (third row) $F_n=8$, (forth row) $F_n=5$ and (fifth row) $F_n=3$. For each chain we plot (from the left the right) (first column) the density of states $\text{DoS}(\nu)$ and (second column) the integrated density of states $\text{iDoS}(\nu)$ as a function of the frequency $\nu$. The black and blue horizontal and vertical lines define the integration boundaries to extract $\text{LDoS}(i,j)$, that are arranged according to the position index $i$ (third column) and conumber index $c(i)$ (fourth column). The vertical axis of the third and fourth column corresponds to the frequency-index $j$.}
\end{figure}

In Fig.~\ref{fig:supp:datatreat_circular} (left column) one can see the measured density of states $\text{DoS}(\nu)= \left< 1 - \Re{S(i,\nu)} \right>_{i}$ for all $F_n=3,5,8,13,21$. Determining the integration borders of each band $j$ is obvious for $F_n=3\text{ and }5$, where the $F_n$ frequency bands are isolated and well separated by clearly visible gaps. While for $F_n=8$ one could eventually still identify 8 bands, although some gaps in between are closing, it becomes impossible for higher $n$ to directly identify all frequency-bands. We therefore calculate the integrated density of states $\text{iDoS}(\nu)=\int^\nu \text{DoS}(\nu') \text{d}\nu'$, which we normalize so that when integrating over all states the $\text{iDoS}(\nu)$ equals $F_n$, the total number of bands ($\int \text{DoS}(\nu') \text{d}\nu'=F_n$). Theoretically in the limit of $\Gamma \rightarrow 0$ and perfectly normalized wave functions, we would obtain a staircase function where we would have $F_n$ big steps with step-height 1, that are comprised of $N_p$ smaller steps, with height $1/N_p$. Since the step corresponding to a single band has a height of 1, one could think of intersecting the $\text{iDoS}(\nu)$ with a set of horizontal lines that have a spacing of 1 in between them. The found intersecting points $\nu^*_k$ ($\text{iDoS}(\nu^*_k)=k,\text{ for all } k \in (0,1,2,...,F_n)$ could then define the integration intervals for each band. 

Due to the non-zero linewidth of our resonances, the $N_p$ smaller steps within a band are completely blurred, while only the plateaus corresponding to the well-visible gaps remain.
Since the antenna coupling $\sigma$ is slightly dependent on the frequency, and the single resonance wavefunctions are slightly overlapping \cite{rei21}, the different states are not properly normalized in the experiment, which translates to slighlty different step heights in the iDoS.
So just intersecting the experimental iDoS with equally spaced lines, does not work very well, as can be seen for the case of $F_n=3$, where the two clearly visible plateaus are not at $\text{iDoS}(\nu)=1$ and $\text{iDoS}(\nu)=2$, as expected if properly normalized, but slightly higher.
We thus use a hybrid approach where we take the frequency-positions of the clearly visible gaps as fixed references and find the frequency-position of the vanished gaps in between by intersecting the iDoS in between with equally spaced lines.
The positions of the visible gaps are extracted by hand and marked as solid black vertical lines in the first two columns of Fig.~\ref{fig:supp:datatreat_circular}.
At the positions where the solid black lines intersect the iDoS, we draw solid blue horizontal lines.
For $F_n=3,5,8$ we were able to identify all gaps, so the solid blue lines divide the iDoS in $F_n$ intervals, but as explained earlier for $F_n=13,21$ not all gaps can be identified.
Whenever we could not identify a gap, we drew additional blue dashed horizontal lines that equally divide the space in between the two solid blue lines by the number of bands that we expected to be in between the clearly visible gaps. To not adjust our results based on our expectations we estimate the number of bands in between two clearly visible gaps (solid blue lines), by rounding the position where the blue lines intersect the iDoS axis to the nearest integer value, and we suppose that this is the number of bands below that gap. In that way, we determine the number of bands in between two solid blue lines. At the frequencies where the dashed blue lines intersect the iDoS, we draw a dashed black vertical line. The black vertical (solid and dashed) lines then define the integration boundaries, which we use to integrate each individual spectrum measured over each resonator, leading to $\text{LDoS}(i,j)$. 

The results can be seen in Fig.~\ref{fig:supp:datatreat_circular}, where $\text{LDoS}(i,j)$ is plotted ordered according to the resonator position indexes $i$ (third column) and to the conumber indexes $c(i)$ (forth column). By averaging $\text{LDoS}(c(i),j)$ over all sites that share the same conumber index, one obtains the smoothed patterns plotted on Fig.~\ref{fig:all_heatmaps}.
Unlike the procedure for linear chains, we do not need to symmetrize our results since this approach allows us to analyze the whole frequency range. For the normalization, the procedure is the same.

\section{Fractal dimensions of the wavefunctions}
\label{sect:boxcount}
We perform a multifractal analysis of the LDoS displayed in Fig.~\ref{fig:heatmap_55}. The fractal dimension $D_q^\psi(j)$ for each state $j$ is defined via an exponential scaling of the generalized inverse participation ratio $\chi^{(n)}(j)$ with the system size $F_n$ (see (\ref{eq:individual fractal dimension})), and the spectrally averaged fractal Dimension $\overline{D_q^\psi}$ is defined by the scaling of the arithmetic average $\left<\chi^{(n)}(j) \right>_j$ over all states in the spectrum.
To investigate the scaling behavior as a function of the system size, one would have to perform the experiment for different system sizes $F_n$, which is impractical in our case, because the maximal possible system size in order to resolve all wavefunctions is 55, which is far from the quasiperiodic limit. Fortunately there is another approach that is commonly used to calculate (fractal) dimensions in various fields of physics and mathematics, namely a box-counting algorithm. The method that we use and present in the following section has already proven itself in the characterization of chaotic systems and multifractal wavefunctions at critical transitions and in quasiperiodic structures \cite{sch91,sch96,chh89,thi13}.

\subsection{Calculation via a box-counting algorithm}

\begin{figure}
 \centering
 \includegraphics[width=\linewidth]{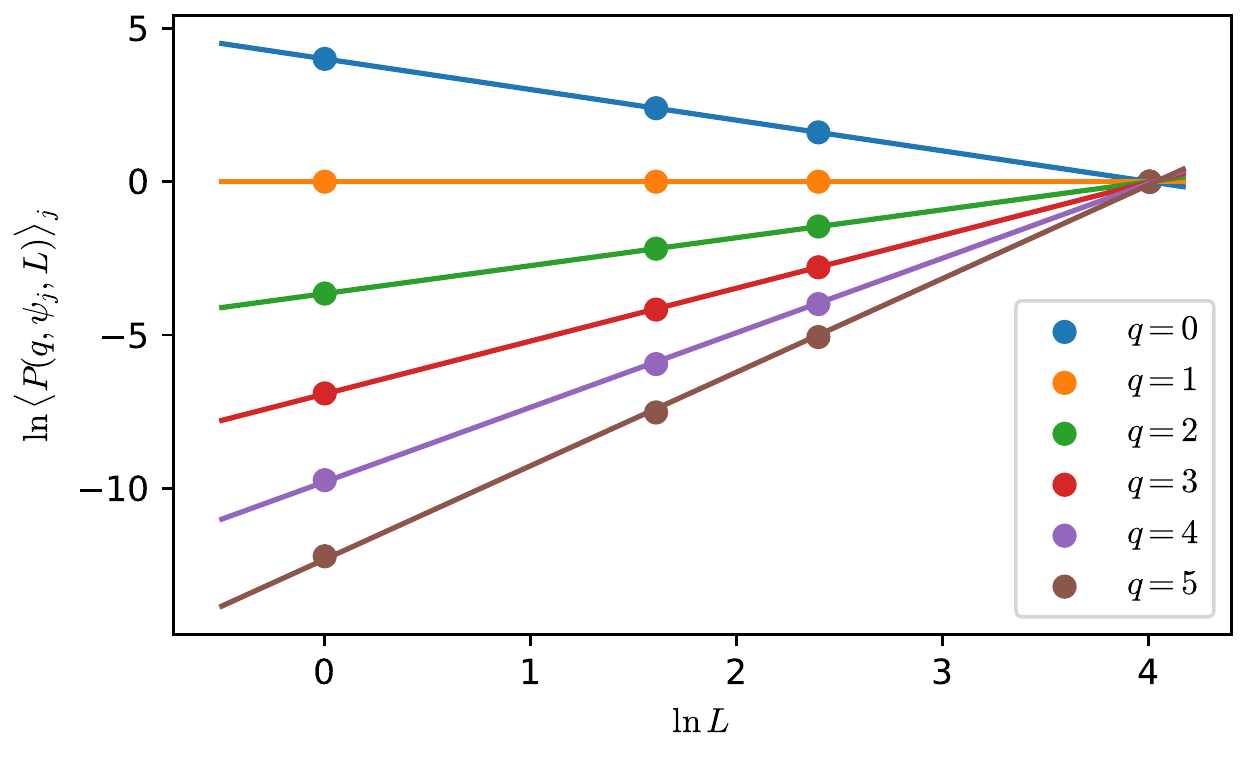}
 \caption{\label{fig:supp:box_counting} Calculated $\ln \left< P(q,\psi_j,L) \right>_j$ for the different box sizes $L$ and some example values of $q$. For each $q$ the data points are fitted individually (solid lines), in order to extract their slope.}
\end{figure}

The main idea behind the box-counting method is to break the system down into small ``boxes" and analyze them individually. By changing the box size and considering smaller and smaller boxes, one can thus deduce scaling properties for the system. We use in the following the notations and formalism presented in reference~\cite{thi13}. We start by dividing our system of size $F_n$ into $B = F_n/L$ boxes of size $L$. Since the system is one dimensional the boxes are actually intervals of length $L$. We then study the spatial distribution of each wavefunction $\psi_j(i)$ by calculating the probability
\begin{equation}
 p_b(\psi_j,L)=\sum_{i \in \text{box } b} | \psi_j(i)|^2
\label{eq:supp:p}
\end{equation}
to find a ``ball" inside box $b$. Repeating this procedure for different box sizes $L$, one can then compute the mass exponent
\begin{equation}
 \tau_q=\lim_{L \to 0} \frac{\ln \left< P(q,\psi_j,L) \right>_j}{\ln L/N}=\lim_{L \to 0} \frac{\ln \left< \sum^B_{b=1} p_b(\psi_j,L)^q \right>_j}{\ln L/N}
\end{equation}
by linear fitting the spectrally averaged quantity $\ln \left< P(q,\psi_j,L)\right>_j$ versus $\ln L$ and extracting the slope. For our system of size $F_n=55$, we consider all box sizes $L = 1,5,11,55$ with integer ration $F_n/L$. In Fig.~\ref{fig:supp:box_counting} we plot and fit $\ln \left< P(q,\psi_j,L) \right>_j$ versus the box size $\ln L$ for some typical values of $q$. We find an excellent agreement between the data points and fit. From the mass exponents $\tau_q$ one can then easily obtain the spectrally averaged fractal dimension $D_q= \tau_q/(q-1)$. 

\subsection{Comparison to numerical results}

In Fig.~\ref{fig:fractal_dimension} one can see the calculated experimental fractal Dimension $\overline{D_q^\psi}$, the theoretical prediction, as well as a confidence interval for our measurement. 
Both the positioning of the resonators as well as the resonance frequency of each resonator have a small variance, which leads to slightly different tight-binding Hamiltonians, wavefunctions, and thus fractal dimensions each time one would perform the experiment. 

The fluctuations of the resonance frequencies have two origins. To place the resonators, we let them drop through a small precision machined down-tube and then apply slight pressure via an plastic rod on top of the dielectric cylinders.
This ensures a good electrical contact between the bottom plate and the resonator, but upon replacing the same resonator several times, the measured resonance frequencies of the very same resonator still vary slightly with a standard deviation of $\approx \SI{0.5}{MHz}$.
Further, the resonators are not identical, resulting in different resonance frequencies as well. Out of a series of 500 resonators, whose resonance frequencies follow approximately a normal distribution with a width of $\SI{40}{MHz}$, we chose the 55 resonators that have the closest resonance frequencies. This results in a difference between the extreme resonance frequencies of $\approx \SI{3}{MHz}$. Since the span of $\SI{3}{MHz}$ is small compared to the width of the distribution of resonance frequencies for the whole series, we suppose that they follow a quasilinear distribution. In addition, we have small variations within the positions upon placing the resonators, which result in slightly varying coupling strength's. In space, these fluctuations are of the order of $\SI{0.05}{mm}$, which induces in the worst case (almost touching resonators) a variation of $4\%$ of the coupling strength.

To estimate the impact of these experimental fluctuations on the extracted fractal dimensions, we simulate the experiment by formulating simple tight-binding Hamiltonians for the 11 different permutations.
We model the resonators' resonance frequencies $\nu^* \sim \SI{7.454}{MHz} + \mathcal{U}(\SI{-1.5}{MHz},\SI{1.5}{MHz}) + \mathcal{N}(0,\sigma_{\nu})$ by employing a uniform distribution with a span of $\SI{3}{MHz}$ combined with a normal distribution with $\sigma_\nu=\SI{0.5}{MHz}$ centered around $\SI{7.454}{GHz}$, which accounts for the variation upon re-placing the same resonator several times.
With $\{ \vec{r}_i \}$ being the exact positions that follow the Fibonacci sequence, we suppose that the actual positions of the resonators $\{ \vec{r}^*_i \}$ follow $\vec{r^*_i}\sim \vec{r_i} + \mathcal{N}(0,\sigma_{\text{pos}})$, supposing a normal distribution with a standard deviation of $\sigma_{\text{pos}}=\SI{0.5}{mm}$ in the $x$ and $y$ directions.
We then calculate the coupling strength between all nearest neighbours $i$ and $k$, by calculating their distances $d_{ik}=|\vec{r^*_i}-\vec{r^*_k}|$ and using the relation $t(d)=\SI{63.2}{MHz} \cdot K_0(\SI{0.481}{mm^{-1}}\cdot d/2) \cdot [ K_2(\SI{0.481}{mm^{-1}} \cdot d/2)+K_0(\SI{0.481}{mm^{-1}} \cdot d/2)] $ between coupling strength $t$ and separation $d$ between two resonators that was extracted from two-resonator measurements~\cite{rei21}.
We diagonalize the Hamitonians in order to find the wavefunctions, average over the different permutations and determine the fractal Dimensions $D_q^*$ via the same box-counting method that we use for the experiment.
We perform this procedure 10000 times, and then for each $q$ the 5th percentile and 95th percentile of the distribution of $D_q^*$ are used as the lower and higher contour line of the gray area in Fig.~\ref{fig:fractal_dimension} respectively, defining a sort of 90\% confidence interval. 

There is a noticeable offset for large values of $q$ between the experimental $\overline{D_q^\psi}$ and the theory.
The fact that the values for one experimental realization have an offset could be explained by experimental fluctuation, since both the experimental points as well as the theoretical curve lie within the confidence interval, but the average values of the simulated $D_q^*$ (white dashed line within the gray area in Fig.~\ref{fig:fractal_dimension}) show an offset compared to the theoretical curve as well.
Next-neighbour couplings within the actual experiment and the way we average over the 11 permutations certainly play a small role, but this offset mainly arises form the finite system size of $F_n=55$. 

\begin{figure}
 \centering
 \includegraphics[width=\linewidth]{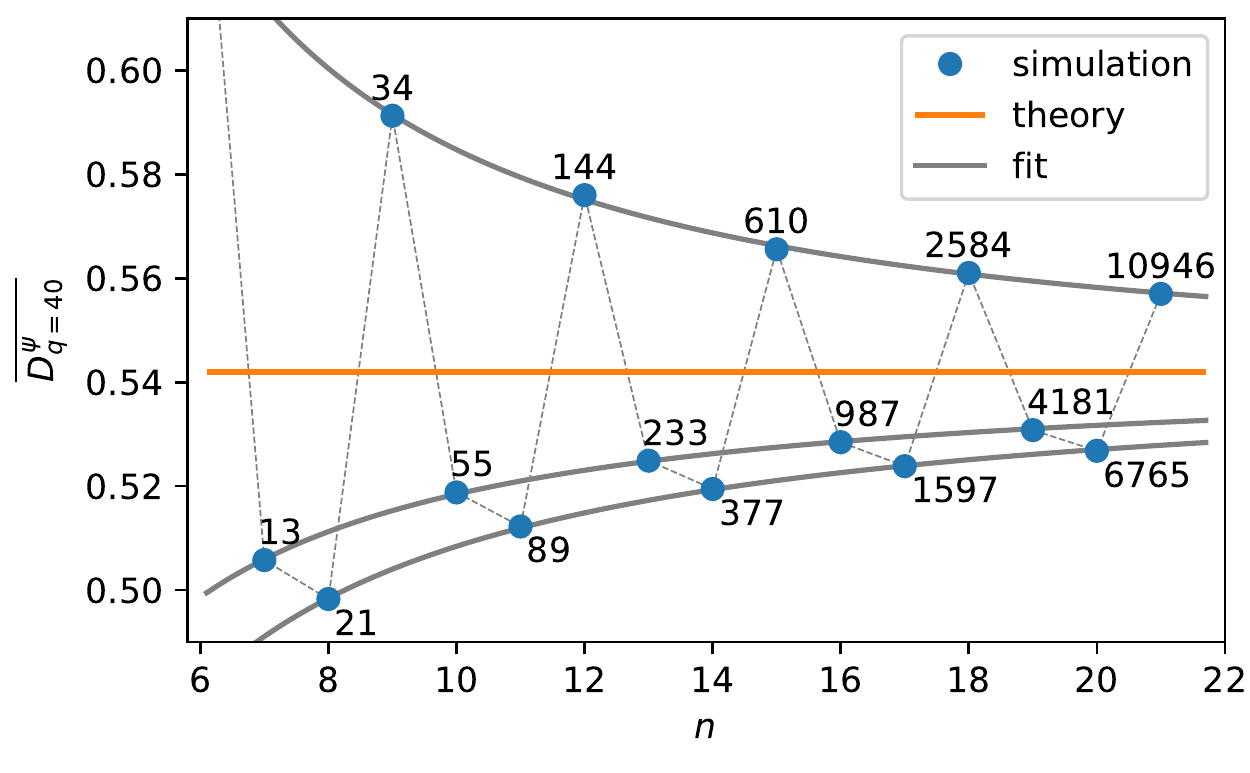}
 \caption{\label{fig:supp:offset_dimension}
 Simulated spectrally averaged fractal dimension $\overline{D_{q=40}^\psi}$ for a high value of $q=40$ as a function of the iteration index $n$ (blue points). The numbers near the blue points are the system size/motif length $F_n$ for each iteration. The orange line presents the theoretical value of the fractal Dimension of the quasiperiodic system $\overline{D_{q=40}^\text{theo}}$. The gray solid lines are fits of the form $D(n)=An^B-y_0$.}
\end{figure}

In Fig.~\ref{fig:supp:offset_dimension} one can see calculated fractal dimensions $\overline{D_{q=40}^\psi}$ for different system sizes $F_n$ for a high value of $q=40$ and $\rho=0.64$.
Since here we only want to compare the effect of the system size, we simulate the only-nearest-neighbor tight-binding system of the $n$th periodic approximation of infinite size by formulating a closed chain of $F_n$ resonators (one single motif) and $F_n$ couplings where we vary over the phase of the connecting coupling between the first and last resonator to account for the periodicity.
One can see that the fractal dimensions converge to the theoretical value in the quasi periodic limit $n\gg 1$, with an oscillating behavior.
A quick explanation of this feature can be given, when looking into the central/most localized state. If the system size is an uneven number, the central state is localized only at the central site, while for an even system size, the central state is localized at the two central positions of $c(i)$. It is thus less localized and therefore it has a greater fractal dimension. Since numerically it is very costly to diagonalize very large matrices, we stop ourselves at a system size of $F_{21}=10946$, which still has a noticeable offset compared to the theoretical value. Then in order to verify that the values converge to the theoretical one, we fit the apparent three different subsets with an algebraic decay $D(n)=An^B-y_0$, where we suppose the same exponent $B$ and offset $y_0$ for all subsets but with different amplitudes $A$. We find $B=-1.298$ and $y_0=0.540$, which corresponds reasonably well to the theoretical value $\overline{D_{q=40}^\text{theo}}=0.542$, considering that the theory was formulated in the strong modulation regime $\rho \ll 1$, where we are far off with $\rho=0.64$.

\section{Renormalization factors \texorpdfstring{$\lambda(\rho)$}{λ(ρ) } and \texorpdfstring{$\overline{\lambda}(\rho)$}{λ(ρ)}}
\label{sect:scaling}
Within the renormalization theory (for the weak coupling dominant case), on obtains a direct recursive construction law for the $\text{LDoS}$ (see eq.~(\ref{eq:renormalization_atomic}) and (\ref{eq:renormalization_molecular})), where the two renormalization factors
\begin{equation}
\overline{\lambda}(\rho)=\frac{2}{(1 + \rho^2)^2 + \sqrt{(1 + \rho^2)^4 + 4\rho^4}}
\end{equation}
and
\begin{equation}
\lambda(\rho)=\frac{1}{1 + \rho^2 \gamma(\rho) + \sqrt{1 + (\rho^2 \gamma(\rho))^2}}
\end{equation}
with $\gamma(\rho)=1/1(1+\rho^2)$, are both dependent on the ratio of the coupling strength $\rho$ \cite{mac16}.

To experimentally estimate the renormalization factors, we make use of the fact that the sum over all states and positions of the $\text{LDoS}$ for a system with motif length $F_n$, sums up to $F_n$ because each of the $F_n$ states was properly normalized $[\sum_i |\psi_j(i)|^2=1$]. 

\begin{figure}
 \centering
 \includegraphics[width=\linewidth]{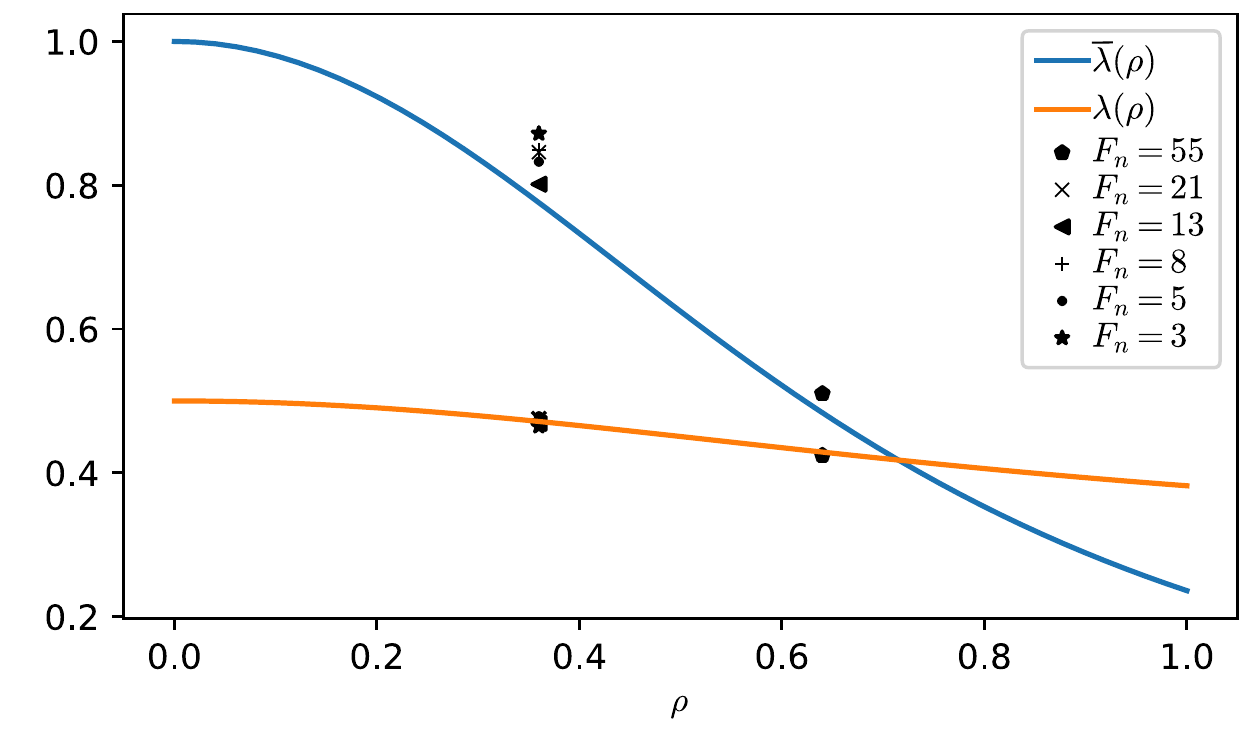}
 \caption{\label{fig:supp:lambda}
 The theoretical renormalization factors $\overline{\lambda}$ (blue line) and ${\lambda}$ (orange line) as a function of $\rho$. The experimentally extracted renormalization factors for the different motif lengths $F_n$ are plotted at their corresponding $\rho$ with different black symbols. 
 }
\end{figure}

We can thus sum up all the pixels that contribute to the central square of the LDoS (corresponding to atomic sites and states) and divide them by the side length of that square (i.e. the number of atomic sites within the chain) to find the renormalization factor $\overline{\lambda}$ for the atomic sites/states. For the molecular sites/states we proceed in the same way, but additionally we average over the four corner squares (corresponding to molecular sites and states).

Since we chose to use different couplings for the linear chains and the circular chains, we can experimentally invest the renormalization factor for two different values of $\rho$: $\rho=0.64$ for the linear chains of $F_n=55$ resonators and $\rho=0.37$ for the circular chains with a smaller motif length. In Fig.~\ref{fig:supp:lambda} one can see the two theoretical curves for $\lambda(\rho)$ and $\overline{\lambda}(\rho)$ as a function of $\rho$, which we compare to the experimentally extracted values (black symbols). 
The experimentally extracted renormalization factors $\lambda(\rho)$ and $\overline{\lambda}(\rho)$ correspond reasonably well with their theoretical prediction, although the extracted  $\lambda(\rho)$ for the molecular sites varies for the different motif lengths $F_n$ and generally shows a slight offset. This can be explained by the small system sizes $F_n$, since the theoretical predictions were formulated in the quasiperiodic limit.

\section{Interchanged couplings}
\label{sect:interchange}
As mentioned in the main text, although we mainly focus our quantitative analyses on the common case of $\rho<1$, we also experimentally investigate the system with interchanged couplings: $t_B$ is now the weaker coupling, and $t_A$ the stronger, $\rho>1$, thus the strong coupling dominates. 
As for the system with $\rho<1$, we investigate the large system ($F_n=55$) by averaging over the 21 different permutations that meet the constraints, and smaller systems by means of circular chains. Note that over the 21 possible permutations, 10 of them are actually mirrored sequences of the others. Since they are experimentally equivalent, we measure only the 11 different permutations that are not mirrored sequences of each other, but average over all 21 permutations by inverting the position axis for the mirrored ones. 

\begin{figure}
    \centering
    \includegraphics[width=\linewidth]{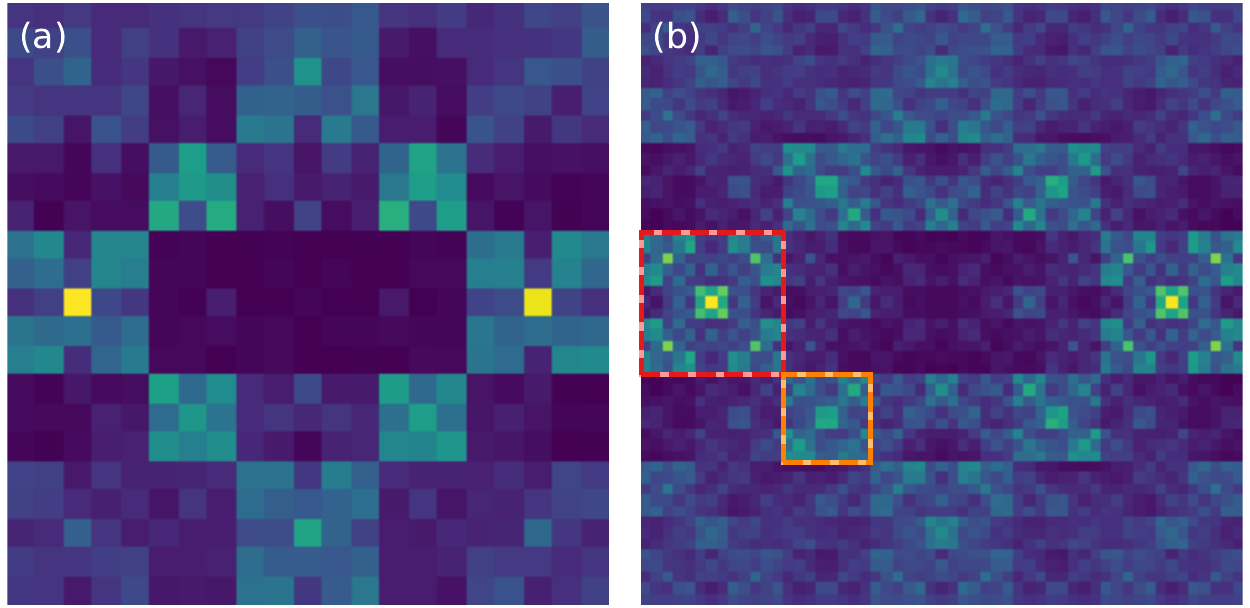}
    \caption{\label{fig:supp:strong_dominant}
    LDoS for motif length $F_8=21$ (a) and $F_{10}=55$ (b) for $\rho<1$. The $x$-axis corresponds to the conumber index $c(i)$ and the $y$-axis to the frequency index $j$ and the same colormap as in Fig.~\ref{fig:heatmap_55} is used. The red and orange square highlight the basic motifs.
    }
\end{figure}

The averaged LDoSs can be seen in Fig.~\ref{fig:supp:strong_dominant}. Instead of single atoms and dimers, as for $\rho<1$, the chains are now composed by dimers and trimers. This results in a different renormalization scheme upon the first deflation of the chain. The effective bond couplings between two neighboring trimers take on only two possible values, arranged according to a Fibonacci sequence but with inverted strong and weak couplings.
One thus passes from the chain $C_n$ to the chain $C_{n-3}$ with $\rho \rightarrow 1/\rho$ when appropriately renormalizing their couplings. In the same way one passes from the chain $C_n$ to the chain $C_{n-4}$ again with $\rho \rightarrow 1/\rho$ for the dimers. All further deflation steps then follow the renormalization laws for $\rho<1$ ~\cite{niu90}.
This explains why the general structure in Fig.~\ref{fig:supp:strong_dominant} is quite different, but we find the same basic motifs as in Fig.~\ref{fig:all_heatmaps}. The red square in Fig.~\ref{fig:supp:strong_dominant}(b) highlights the basic motif associated with the trimers, which can be found in Fig.~\ref{fig:all_heatmaps}(d,f), while the orange square highlights the basic motif associated with the dimers, which again can be found in Fig.~\ref{fig:all_heatmaps}(c).

\end{document}